\documentclass[a4paper, 11]{JHEP3}

\usepackage[english]{babel}
\usepackage{amsmath,amssymb,amsbsy,amstext}
\usepackage{graphicx}
\usepackage{amsfonts}

\usepackage{mathrsfs}
\usepackage{subfigure}
\usepackage{epsfig,latexsym}

\usepackage{cite}
\usepackage{graphicx}

\def\baray{\begin{eqnarray}}
\def\earay{\end{eqnarray}}
\def\ba{\begin{eqnarray}}
\def\ea{\end{eqnarray}}

\newcommand{\be}{\begin{eqnarray}}

\newcommand{\ee}{\end{eqnarray}}

\newcommand{\dnm}{\Delta N_{\rm max}}
\newcommand{\fb}{\phi_{b}}

\newcommand{\Mpl}{M_{p}}
\newcommand{\ab}{\vec A_{b}}
\newcommand{\ex}[1]{\langle #1 \rangle}
\newcommand{\edb}{\vec E \cdot \vec B}
\newcommand{\vf}{\varphi}
\newcommand{\va}{\vec A}

\newcommand{\R}{\mathcal{R}}
\newcommand{\fid}{f^{ id}_{NL}}
\newcommand{\xic}{\xi_{\ast}}
\newcommand{\Dsr}{\Delta_{\R,{\rm sr}}^{2}(k)}
\newcommand{\Dsrki}{\Delta_{\R,{\rm sr}}^{2}(k_i)}
\newcommand{\floc}{f_{NL}^{\rm loc}}
\newcommand{\dep}{\delta \phi}
\newcommand{\es}{\epsilon_{\ast}}
\newcommand{\sr}{\mathcal{O}(\es,\eta_{\ast})}
\newcommand{\cq}{C_{k}}

\title{Observational Constraints on Gauge Field Production in Axion Inflation}

\author{P.~D.~Meerburg $^{1}$ and E.~Pajer$^{2}$\\
$^1$ Department of Astrophysical Sciences, Princeton University, Princeton, NJ 08540 USA\\
$^2$ Department of Physics, Princeton University, Princeton, NJ 08544 USA \\}

\date{\today}

\abstract {Models of axion inflation are particularly interesting since they provide a natural justification for the flatness of the potential over a super-Planckian distance, namely the approximate shift-symmetry of the inflaton. In addition, most of the observational consequences are directly related to this symmetry and hence are correlated. Large tensor modes can be accompanied by the observable effects of a the shift-symmetric coupling $\phi F\tilde F$ to a gauge field. During inflation this coupling leads to a copious production of gauge quanta and consequently a very distinct modification of the primordial curvature perturbations. In this work we compare these predictions with observations. We find that the leading constraint on the model comes from the CMB power spectrum when considering both WMAP 7-year and ACT data. The bispectrum generated by the non-Gaussian inverse-decay of the gauge field leads to a comparable but slightly weaker constraint. There is also a constraint from $\mu$-distortion using TRIS plus COBE/FIRAS data, but it is much weaker. Finally we comment on a generalization of the model to massive gauge fields. When the mass is generated by some light Higgs field, observably large local non-Gaussianity can be produced.}


\begin{document}

\maketitle

\section{Introduction}

Inflation is a UV sensitive mechanism in that it drastically depends on the size of Planck suppressed operators up to dimension six or more. One of the main objectives of inflationary model building is to find natural constructions in which these operators have values compatible with current observations. If we now further focus on inflationary models that can produce observable tensor modes, in which the scale of inflation is close to the GUT scale, the UV sensitivity becomes even more pungent: there is an infinite number of higher dimensional operators that need to be small in order for inflation to take place at all. Facing these obstacles, the use of symmetries constitutes a very promising way to construct successful and robust models. 

Axions enjoying a shift symmetry to all orders in perturbation theory have been recognized long ago \cite{natural} as promising inflaton candidates. Since then many different constructions have been proposed \cite{models,monodromy}. Without getting into the very interesting and challenging details of model building, our focus here is on phenomenological consequences that all this class of models have in common. The assumption of a shift symmetry already puts strong constraints on the allowed interactions of the inflaton. One shift-symmetric interaction that has so far proven to be extremely relevant for observation is a $\phi F \tilde F$ coupling of the inflaton to a gauge field. On the one hand, this coupling can be responsible for a non-perturbative periodic correction to the inflaton potential\footnote{The origin of the non-perturbative correction in explicit string constructions is somewhat different, but leads to a similar low energy effective theory.} \cite{monodromy}. The consequences of this oscillatory correction have already been discussed elsewhere \cite{osci,monodromy,Behbahani:2011it} and will not be further discussed in the present work. On the other hand, during inflation the $\phi F \tilde F$ coupling leads also to a window of tachyonic instability for the gauge fields, which are then copiously produced. There is a regime in which these gauge quanta are abundant enough to affect the background dynamics at the time when CMB scales left the horizon as for example in warm inflation \cite{warm,lorenzo}. We will not consider this regime since it requires non-minimal modifications in order to be compatible with the data. Instead, following \cite{BP} we focus on the more conventional regime in which most of the exponential expansion is driven by standard slow-roll inflation and the gauge quanta only affect the perturbations. We allow for some additional e-foldings driven by dissipative effects as long as this takes place at the end of inflation, i.e.~for scales too small to be observable.  

There are several observational channels. Through an inverse decay process two gauge quanta can produce an inflaton perturbation hence affecting all scalar primordial N-point functions \cite{BP} (to be reviewed in section \ref{s:r}). The effect on the power spectrum is a blue-tilted contribution, which grows on smaller scales. Since the inverse decay produces non-Gaussian inflaton perturbations, the bispectrum is non-vanishing and peaks on equilateral configurations. Gauge fields also source tensor modes. The effect is negligible at CMB scales but grows on smaller scales, such that it is observable by gravitational wave interferometers such as Advanced LIGO/Virgo \cite{Sorbo,BPP}.

%%%%%%%%%%%%%%%%%%%%%%%%%%%%%%%%%%%%%%%%%%%%%%%%%%%%

\subsection{Summary}\label{s:sum}

For the reader excited mostly about our main results rather than the details of the analysis, we commence this paper with a summary. We compare the predictions of axion inflation allowing for a shift-symmetric coupling
\be\label{coup}
L&\supset & -\frac{\alpha}{4f}\phi F_{\mu\nu}\tilde{F}^{\mu\nu},
\ee
with observations. Here $f$ is the axion decay constant with dimensions of mass and $\alpha$ is the dimensionless size of the coupling. Let us stress that in order to have successful reheating the inflaton should couple to something. The particular coupling in \eqref{coup} is a natural one to consider since it is allowed by all the symmetries of the theory.

\begin{table}%EP
\centering
\begin{tabular}{ |c|c|c|c|c|c|} 
\hline\hline	
  & COBE/FIRAS & WMAP7 $f_{NL}$ & WMAP7 & WMAP7+ACT &\tabularnewline
   \hline
flat prior &  $\xic < 3.6 $ & $\xic < 2.45 $ &$\xic < 2.66 $  &$\xic < 2.41 $  &\tabularnewline
 \hline
log prior &	$\xic < 3.2$ &$\xic < 2.22 $& $\xic < 2.51 $& $\xic < 2.15  $&	\tabularnewline
 \hline 
 \end{tabular} 
 \caption{The table summarizes the various $95\%$ CL constraints we derived in this work for the quadratic model of section \ref{s:sum}. }
 \label{tab:analysis1} 
 \end{table}

\begin{table}
\centering
\begin{tabular}{|c|c|c|c|} 
\hline\hline	
  & WMAP7 & WMAP7+ACT &\tabularnewline
   \hline
flat prior & $\xi_* < 2.5 $  &$\xi_* < 2.5 $  &\tabularnewline
 \hline
log prior & $\xi_* < 2.18 $& $\xi_* < 2.14 $&	\tabularnewline
 \hline 
 \end{tabular} 
 \caption{The table summarizes the various $95\%$ CL constraints we derived in this work for the generic model of section \ref{s:sum}. The flat prior is both less theoretically motivated and gives rise to convergence issues discussed in section \ref{s:an}, so one should mainly focus on the results for the log-flat prior, as far as the generic model is concerned. }
 \label{tab:analysis2} 
 \end{table}

\begin{figure}
\centering
\includegraphics[width=.8\textwidth]{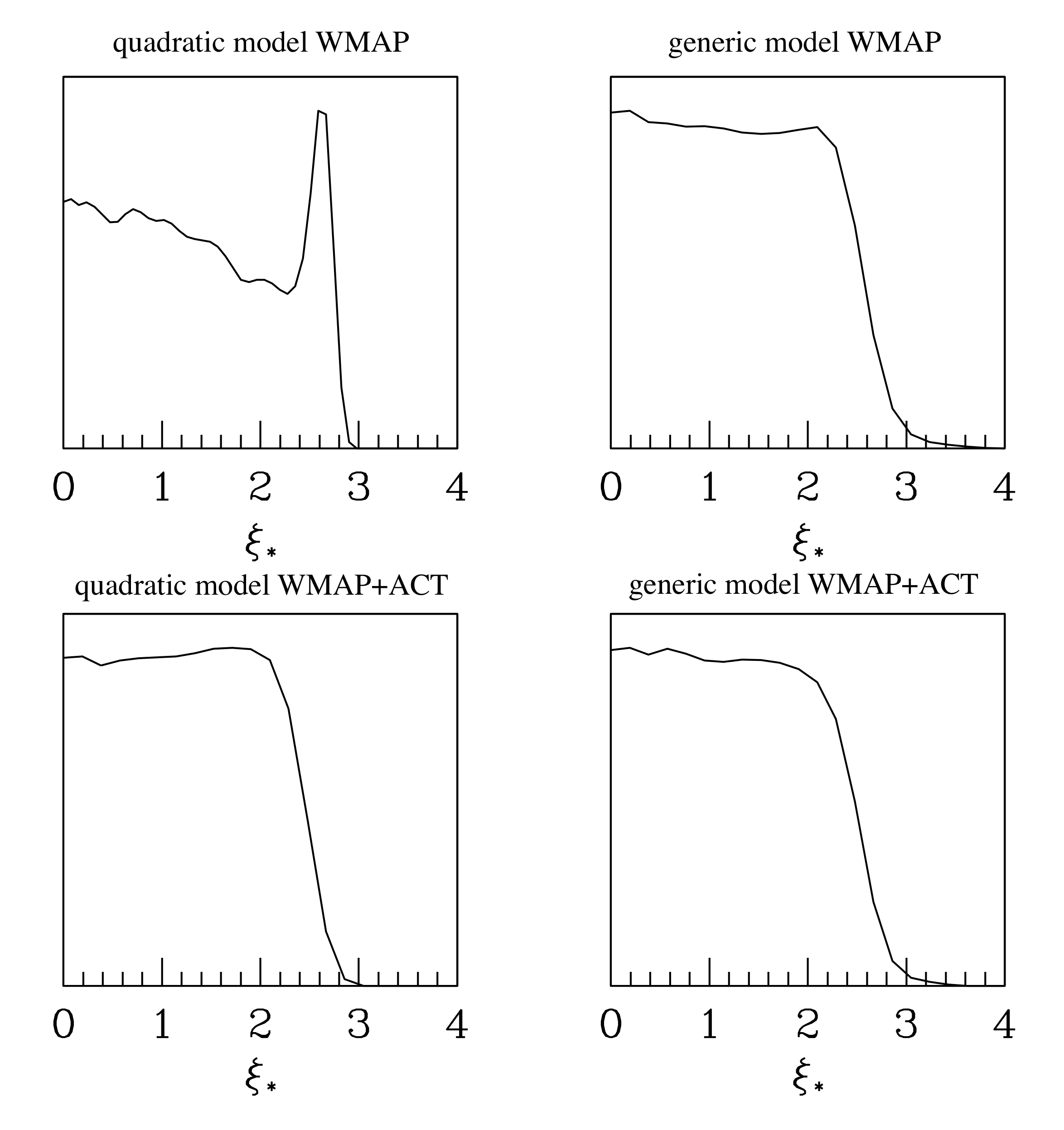}
\caption{The four panels show the marginalized likelihood as function of $\xic$. $\xic\gtrsim 2.5$ is ruled out so the likelihood goes quickly to zero, while $\xic\lesssim 2$ has no appreciable effect on the power spectrum such that the likelihood becomes constant. \label{fig:marg}}%EP
\end{figure}

As mentioned in the introduction, gauge perturbations are copiously produced during inflation due to the time variation of the inflaton $\phi$. We focus on the phenomenological consequences of the inverse decay of gauge quanta ($ A$) into inflaton perturbations ($\delta\phi$). We present our final results as constraints on the combination $f/(\Mpl \alpha)$ or equivalently on the parameter
\be\label{xi}
\xi(k)\equiv \frac{\dot \phi  \alpha}{2H  f}=\sqrt{\frac{\epsilon}{2}} \frac{\alpha}{f}\,.
\ee
For concreteness we quote the bounds on $\xic\equiv\xi(k_{\ast})$, where $k_{\ast}=0.002 \,{\rm Mpc}^{-1}$ is our choice of the pivot scale. We consider two different theoretical priors. In both cases the choice of the upper bound is mostly irrelevant\footnote{The upper bound is relevant if one wants to compute the Bayesian \textit{evidence}. In all cases we consider, the evidence for a model containing the effects of gauge production is smaller than the evidence of the standard $\Lambda$CDM model as can be seen from the marginalized likelihood for $\xic$ in figure \ref{fig:marg}.} as long as it is much larger than $\xic \sim$ few, since the posterior distribution drops to zero extremely fast for $\xic\gtrsim 3$ (see figure \ref{fig:marg}). The two priors we consider are:
\paragraph{Flat prior:} Constant prior on $\xic$ in the interval $\{0-10\}$. This is more practical for the data analysis but is theoretically  less motivated. In terms of physical parameters this choice corresponds to a flat prior on the axion decay constant, but we have no reason to believe that this should be fixed at some precise scale.
\paragraph{Log-flat prior:} Log-constant prior on $\xic$ in the interval $\{10^{-1}-10^{2}\}$. We feel that this choice is better motivated theoretically. It corresponds to having some fixed $\alpha\sim \mathcal{O}(1)$ and a log-flat distributed axion decay constant in the interval\footnote{The factor of $2/3$ is chosen to get a nicer interval in $\xic$.} $(2/3) \times 10^{-3}<f/\Mpl<2/3$. The lower (upper) bound of this $\xic$ ($f/\Mpl$) interval is motivated by the fact that super-Planckian axion decay constants seem hard to obtain in UV finite theories of gravity \cite{Banks:2003sx}. The lower bound is somewhat arbitrary.%The value $f/\Mpl>10^{-4}$ has been chosen so that $f$ is always much larger than the Hubble constant during inflation.

For the comparison with the data we take the power spectrum to be \cite{BP}
\be\label{ps}
\Delta_{\R}^{2}(k)&= &\Dsr \left[1+\Dsr\,f_{2}(\xi)\,e^{4\pi\xi}\right]\,,
\ee
where 
\be
\ex{\R(\vec k)\R(\vec k')}&\equiv&(2\pi)^{3} P_{\R}(k)\delta^{3}\left(\vec k+\vec k'\right)\equiv (2\pi)^{3} \frac{2 \pi^{2}\Delta_{\R}(k)^{2}}{k^{3}}\delta^{3}\left(\vec k+\vec k'\right)\,,\\
\Dsr&\equiv &\Delta_{\R}^{2}(k)|_{\xi=0}=\left(\frac{H}{2\pi}\right)^{2} \frac{H^{2}}{|\dot \phi|^{2}}=\frac{H^{4}_{\ast}}{(2\pi)^{2}|\dot \phi_{\ast}|^{2}} \left(\frac{k}{k_{\ast}}\right)^{n_{s}-1}\,,\label{del}
\ee
and $f_{2}(\xi)$ is defined in \eqref{ff2} and is handled numerically in the data analysis. Here all time dependent quantities should be evaluated at horizon crossing of the relevant mode and a star refers to the pivot scale $k_{\ast}=0.002\, {\rm Mpc}^{-1}$. The slow-roll expansion gives a good approximation already at linear order
\be 
n_s - 1 &=& -2\epsilon_* - \eta_* \simeq 6\epsilon_V + 2 \eta_V\,,\\
\xi & =&\xi_{\ast} \left[1+\frac{\eta_{\ast}}{2}\log \left(\frac{k}{k_*}\right)\right]+\mathcal{O}(\epsilon^{2})\,,\label{xifscale}
\ee
where
\be
\epsilon\equiv -\dot H/ H^{2}\,,
\;\;\;\;\;\;\;\;\;\;\;\;\;\;\; \eta\equiv \dot \epsilon/(\epsilon H)\,,
\ee
while potential slow-roll parameters are defined by 
\be
\epsilon_{V}\equiv\frac{ \Mpl^{2}}{2} \left(\frac{V'}{V}\right)^{2}\,, \quad \eta_{V}\equiv \Mpl^{2} \frac{V''}{V}\,.
\ee
In order to numerically evaluate \eqref{ps} we need to specify an inflaton potential. In the rest of our analysis we consider two models:

\paragraph{Quadratic Model:} We assume a quadratic inflaton potential $V(\phi)=m^{2}\phi^{2}/2$ and fix the number of e-foldings $N$ between the end of inflation and when the pivot scale $k_{\ast}$ leaves the horizon to be $N=60$. We can then solve the background homogeneous equations of motion and compute $H,\,\phi(t)$ and $\xi(t)$. This fixes the slow-roll parameters in \eqref{xifscale}. Then the primordial power spectrum \eqref{ps} contains only \textit{two parameters}:  $\{ \xic,\,\Delta_{\R,{\rm sr}}^{2}(k_{*}) \} $.

%This fixes the running and allows us to solve the background in order to compute $\Delta_{\R}$. In this model, the primordial parameters to be constrained are $A_s$ (the power of the primordial power spectrum) and $\xic$. The function $f_2$ is computed numerically, and the modified version of the code reads in a sufficient large array, consisting of 3 columns: $\xi$, $k$ and $\Delta_{\R}$. As explained, the scale dependence is due to a changing background for each mode through $H$ and $\xi$, which itself is a function of $H$.

\paragraph{Generic Model: } Instead of specifying a potential, we can vary the slow-roll parameters in the fit. The two additional parameters can be chosen to be $\{\epsilon_{\ast},\eta_{\ast}\}$ or equivalently $\{n_{s},r\}$ in \eqref{xifscale}. In this model the primordial power spectrum \eqref{ps} is described by \textit{four parameters}: $\{\xic,\,\Delta_{\R,{\rm sr}}^{2}(k_{\ast}),\epsilon_{\ast},\eta_{\ast}\}$.\\

Currently the strongest bound on the inverse decay phenomenology comes from WMAP7 \cite{Komatsu:2010fb} plus ACT \cite{Dunkley:2010ge} data (see section \ref{s:an} for details). For the quadratic model this is $\xic<2.15$ or equivalently $f/\alpha>0.032 M_p $ at $95\%$ CL for a log-flat prior. For a flat prior the same data leads to $\xic< 2.41$ or equivalently $f/\alpha>0.035 M_p$ at $95\%$ CL. The model independent constraints are remarkably similar. As originally derived in \cite{BP}, the current bound coming from non-Gaussianity can be estimated from the analysis of the equilateral template (with which the inverse decay template has a cosine of $0.94$). As discussed in section \ref{ss:bi}, we find $\xic<2.45$ and $\xic<2.22$ at $95\%$ CL for a flat and log-flat prior, respectively. We also consider CMB spectral distortion of the $\mu$-type and find that using the best current data from COBE/FIRAS leads to a constraint which is much weaker than those from temperature anisotropies. For convenience we have collected all the bounds in tables \ref{tab:analysis1} and \ref{tab:analysis2}. The forecast constraints from Planck and ACTPol are summarized in table \ref{tab:analysis3}.

Our analysis shows that in the region of parameter space allowed by constraints on the scalar power spectrum, non-Gaussianity from inverse decay is currently unobservable. On the other hand, in the case of a detection of an upward bent of the power spectrum at small scales, inverse decay non-Gaussianity would be an important signal to be looked for in the data\footnote{This is somewhat analogous to the results presented in \cite{Behbahani:2011it}, albeit for a different set of phenomenological signatures of axion inflation, where the authors point out that a periodic modulation of the inflationary potential lead to a parametrically larger signal in the power spectrum compared with the bispectrum.}.
%EP

Finally in section \ref{s:h}, we consider a generalization of the model in which we allow for the gauge field $A$ to get a mass from the vev of an additional Higgs-like field $h$. Quantum oscillations of this extra field can convert into adiabatic curvature perturbations. The phenomenologically interesting (but technically unnatural) regime is when $h$ is light as compared to Hubble. In this case large non-Gaussianity of the local type can be  produced. The constraint on $\xi_*$ then depends on a combination of the vev of $h$, the scale of inflation and the gauge coupling.

%%%%%%%%%%%%%%%%%%%%%%%%%%%%%%%%%%%%%%%%%%%%%%%%%%

\section{Review of inverse decay in axion inflation}\label{s:r}

In the section we review how the presence of the coupling \eqref{coup} affects the curvature perturbations. We focus on the results relevant for our analysis and refer the reader to \cite{BP,Sorbo} and appendices \ref{a:run} and \ref{a:bi} for further details on the derivation. We consider the effective Lagrangian
\be\label{L}
S=-\int d^{4}x \sqrt{-g}\left[\frac12R+\frac12 \partial \phi^{2}+\frac14 F^{2}+\alpha\frac{\phi}{4f}F\tilde F+V(\phi)\right]\
\ee
where the inflaton $\phi$ enjoys a shift-symmetry to all orders in perturbation theory. We assume that the gauge coupling is such that non-perturbative effects are negligibly small and refer the reader to \cite{monodromy,osci,Behbahani:2011it} for the study of a different regime. Assuming a homogeneous and slowly evolving background, one can solve the equations of motion for $A$. This solution can be used to compute the contribution of the gauge field to the homogeneous equations of motion
\be
\ddot{\phi} + 3 H \dot{\phi} + V' &=& \frac{\alpha}{f} \langle \vec{E} \cdot \vec{B} \rangle \, , \label{fh}\\
3 H^2 M_p^2 &=& \frac{1}{2} \dot{\phi}^2 + V + \frac{1}{2} \langle \vec{E}^2 + \vec{B}^2 \rangle \label{Fried}\, . 
\ee
where the two expectation values on the right-hand side can be estimated to be
\be
\langle \vec{E} \cdot \vec{B} \rangle  \simeq - 2.4 \cdot 10^{-4}   \, \frac{H^4}{\xi^4} \, {\rm e}^{2 \pi \xi} \;\;,\;\; \langle \frac{\vec{E}^2+\vec{B}^2}{2} \rangle \simeq 1.4 \cdot 10^{-4} \frac{H^4}{\xi^3} {\rm e}^{2 \pi \xi} \, ,
\ee 
with $\xi\equiv\alpha \dot \phi/(2fH)$. One can check that the contribution to the $\phi$ equation of motion \eqref{fh} is always more important than the one to the Friedman equation \eqref{Fried}. Given that $\xi\propto \sqrt{2\epsilon}$, it generically grows as inflation proceeds. When $\xi\gtrsim 5$ the gauge field starts back-reacting sizably on the homogeneous evolution. Here we consider the regime in which $\xi\lesssim3$ when the observable CMB scales left the horizon, in which case the effects of the gauge fields are relevant only for the perturbations. In general $\xi$ will grow large towards the end of inflation, acting as an effective friction term for $\phi$. This generates additional e-foldings that we consistently take into account by solving  the equations of motion numerically as explained in \cite{BPP}. 

To compute the primordial curvature perturbations one needs to solve a linearized equation for the $\delta\phi$ perturbations. Schematically one finds $\delta\phi=\delta \phi_{sr}+\delta \phi_{id}$, where the former contribution is the one of standard for slow-roll inflation while the latter is generated by the inverse decay of gauge field perturbations. Converting to curvature perturbations $\R$ (see \eqref{Rtof}), one finds the power spectrum \cite{BP}
\be
\langle \R(\vec k)\R(\vec k')\rangle&=&(2\pi)^{3} \delta \left(\vec k+\vec k'\right)P_{\R}(k)=(2\pi)^{3} \delta \left(\vec k+\vec k'\right) \frac{2\pi^{2}\Delta_{\R}^{2}(k)}{k^{3}}\,,\label{Pspec}\\
\Delta_{\R}^{2}(k)&= &\Dsr \left[1+\Dsr\,f_{2}(\xi)\,e^{4\pi\xi}\right]\,, \label{Delta}
\ee
where $f_{2}$ is defined and discussed around \eqref{ff2} and 
\be
\Dsr\equiv \frac{H^{4}}{(2\pi)^{2}|\dot \phi|^{2}}=\frac{H^{4}_{\ast}}{(2\pi)^{2}|\dot \phi_{\ast}|^{2}} \left(\frac{k}{k_{\ast}}\right)^{n_{s}-1}
\ee
is the power spectrum in the absence of a coupling to gauge fields. We remind the reader that here and in the rest of the paper we denote with a star a time-dependent quantity, such as\footnote{$\tau$ is conformal time defined by $a\,d\tau=dt$.} $H(\tau)$, $\dot \phi(\tau)$ or $\epsilon(\tau)$, evaluated when a certain fixed pivot scale $k_{\ast}$ crosses the horizon. All time dependent quantities without a start should be evaluated at horizon crossing of the relevant mode, i.e.~when $-\tau k=1$. Notice that what is most crucial for the comparison with observations is  the scale dependence of \eqref{Delta}, since the overall amplitude is undetermined in the inflationary mechanism. The largest violation of scale invariance comes typically from the scale dependence of $\xi$, given in \eqref{xifscale}. For the suspicious reader we have collected in appendix \ref{a:run} some computational details leading to $\eqref{Delta}$.

One can also compute the bispectrum \cite{BP}. As argued in appendix \ref{a:bi} we will consider
\be
\langle \R^{3} \rangle &=& \frac{3}{10}(2\pi)^7 \left[\prod_{i=1}^{3} e^{2 \pi \xi(k_{i})} \Dsrki \right] \delta(\sum \vec{k}_i) \frac{\sum k_i^3 }{\prod k_i^3}f_3(\xi,k_2/k_1,k_3/k_1).
\label{eq:bispectrum}
\ee
$f_3(\xi,k_2/k_1,k_3/k_1)$ is a function similar to $f_2$ first derived in \cite{BP} and the full expression can be found in appendix \ref{a:bi}. %We find that slow-roll corrections only have limited effect on the final constraints. 

%%%%%%%%%%%%%%%%%%%%%%%%%%%%%%%%%%%%%%%%%%

\section{Constraints from spectral distortion}%mu

In \cite{Hu:1994bz} it was pointed out that bounds on (or measurements of) the $\mu$-type spectral distortion of the CMB provide a direct bound on (or measurement of) the log-integral of the primordial scalar power spectrum in the approximate range $50\lesssim k\times {\rm Mpc} \lesssim10^{4}$. Recently this and other aspects of spectral distortion have experienced a revival \cite{munew,Chluba:2012we,Chluba:2012gq,Khatri:2011aj,Pajer:2012vz}. Some useful older references are \cite{mu,ddz}. Since the inverse decay of gauge fields into scalar perturbations leads to a blue tilt in the primordial power spectrum it is natural to ask what the constraint from $\mu$-distortion is\footnote{Close to the completion of our work, \cite{Chluba:2012we} appeared where analogous constraints are computed for a few other classes of models.}. We show in the following that the current constraints (from TRIS \cite{tris} COBE/FIRAS \cite{cobe}) are much weaker than those from the CMB power spectrum and bispectrum, as we will see in the next section. For this reason we limit the analysis to the quadratic model. The main conclusion similarly holds for a generic model.

%%%%%%%%%%%%%%%%%%%%%%%%%%%%%%%%%%%%%%%%%%%%%%%%%%%%%%%

\subsection{Review of $\mu$-type spectral distortion}

We start with a quick review of the relevant CMB physics following \cite{Hu:1994bz,Chluba:2011hw,Khatri:2011aj}. At early times, $z\gg2\times 10^{6}$, the occupation number of photons is extremely well described by a black body formula. As the universe expands and cools down, interactions become less efficient. After $z\simeq 2 \times 10^{6}$ equilibrium relies mostly on elastic Thomson scattering. Since Thomson scattering takes one photon in and gives one photon out, effectively the number of photon becomes frozen. Let us now consider perturbing the system by adding some energy $\delta E$. Thomson scattering redistributes the energy (kinetic equilibrium), but there is a conserved number, so the equilibrium distribution is a Bose-Einstein distribution with a non-vanishing frequency-dependent chemical potential $\mu(\nu)$, rather than a black body spectrum. A Boltzmann equation for $\mu(\nu)$ can be written down (Kompaneets \cite{Kom}) and solved. The result is that $\mu(\nu)$ is almost constant except for very low frequencies, where light photons can be produced at low energy cost. As the expansion of the universe proceeds, even elastic Thomson scattering is not efficient enough after $z\simeq 5 \times 10^{4}$ to redistribute energy and momentum if the system is perturbed. There is a smooth transition to a situation in which an energy injection into the system produces a deformation of the photon spectrum quite different from the one due to $\mu$-distortion, which then can be distinguished observationally. Finally after the photons decouple, around $z\simeq 1100$, any imprinted distortion is maintained and can be observed  in the CMB. 

\begin{figure}
\centering
\includegraphics[width=.8\textwidth]{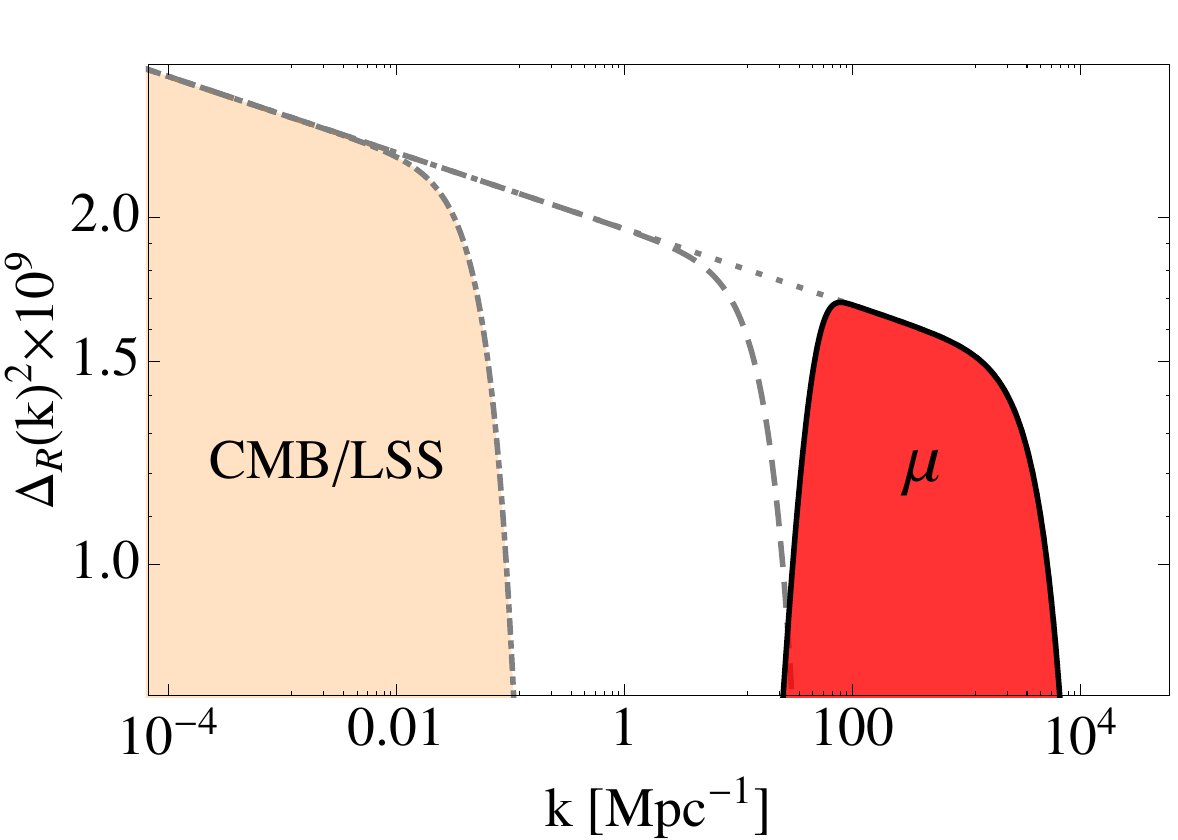}
\caption{The plot shows the primordial scalar power spectrum including Silk damping at three different times. More specifically we plot $\Delta_{\R}^{2}(k) e^{-k^{2}/k_{D}^{2}(z)}$ for $z=1100,\,5\times 10^{4},\,2\times 10^{6}$. Measurements of $\mu$-distortion are sensitive to the red region on the right side, i.e.~much smaller scales than those probed by large scale structures and temperature anisotropies (on the left). Figure taken from \cite{Pajer:2012vz}.}
\end{figure}

The energy injection and associated spectral distortion that we want to consider is the one due to dissipation of adiabatic acoustic waves. As primordial superhorizon perturbations re-enter the horizon during radiation domination they start oscillating and eventually they dissipate part or all of their energy due to diffusion damping \cite{Silk}. The energy that is dissipated and hence injected into the photon-baryon plasma from $z_{i}\equiv2 \times 10^{6}$ to $z_{f}\equiv 5\times 10^{4}$ leads to $\mu$-distortion.

%%%%%%%%%%%%%%%%%%%%%%%%%%%%%%%%%%%%%%%%%%%%%%%%%%%%%%

\subsection{Constraints on inverse decay from $\mu$-distorsion}

Very precise predictions for $\mu$-distortion are best obtained by numerically evolving the Boltzmann equations for the photon-electron-baryon plasma. Here, on the other hand, we will make the necessary approximations in order to be able to treat the problem analytically. We believe that our main conclusions, i.e.~that the bound from $\mu$-distortion is weaker than those from temperature anisotropies, are unaffected by these simplifications. The relations we will use is \cite{Hu:1994bz}
\be\label{mu}
\mu\simeq 3 \int_{k_{D,i}}^{k_{D,f}} d\ln k\,\Delta_{\R}^{2}(k) \left[e^{-k^{2}/k_{D}^{2}(z)}\right]^{z_{f}}_{z_{i}}\,,
\ee
where $k_{D}(z)\simeq 4\times 10^{-6} z^{3/2}\,{\rm Mpc}^{-1}$ is the damping scale. As we reviewed in section \ref{s:r}, the primordial scalar power spectrum in models of axion inflation accounting for inverse decay of gauge fields is given by \eqref{Delta}.
%\be
%\Delta_{\R}^{2}(k)\equiv \frac{H^{4}}{(2\pi)^{2}|\dot \phi|^{2}} \left[1+\frac{H^{4}}{(2\pi)^{2}|\dot \phi|^{2}}\,f_{2}(\xi)\,e^{4\pi\xi}\right]\,.
%\ee
%where we remind the reader that $\xi=\dot \phi \alpha/(2 H f)$, with $f$ the axion decay constant determining the coupling $\phi F\tilde F/f$.

\begin{figure}
\centering
\includegraphics[width=.6\textwidth]{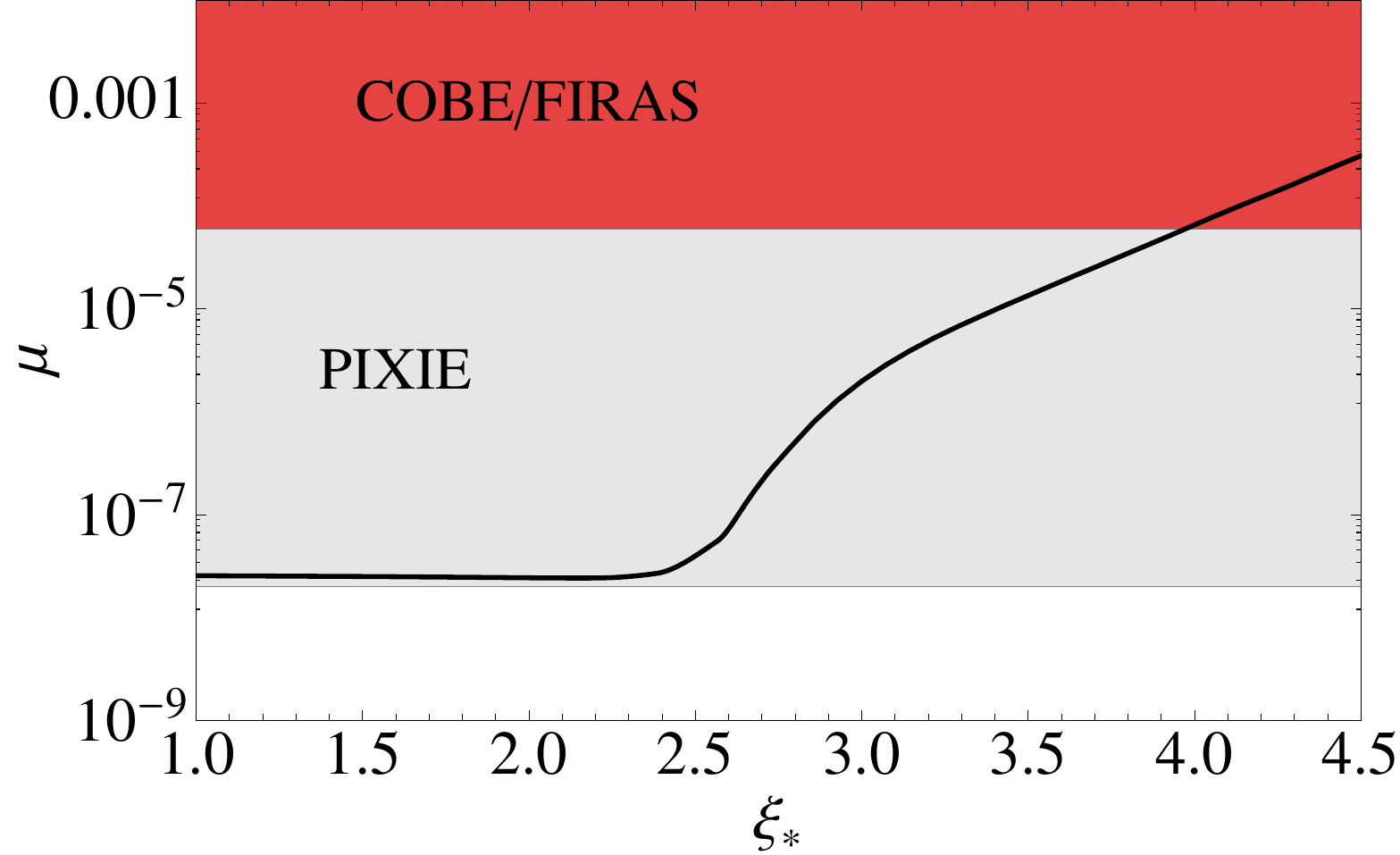}
\caption{The plot shows $\mu$ as function of $\xic$. $\mu<6\times 10^{-5}$ is the $95\%$ CL exclusion contours from the combination of TRIS \cite{tris} and COBE/FIRAS \cite{cobe}, while from forecasts of the PIXIE experiment \cite{Kogut:2011xw} one gets $\mu<2 \times 10^{-8}$ at $95\%$ CL.\label{f2}}
\end{figure}

Using \eqref{Delta} and \eqref{mu} we derive the dependence of $\mu$-distortion on $\xic$ for the quadratic model of section \ref{s:sum}. This is plotted in figure \ref{f2} together with the most recent constraint from TRIS \cite{tris} plus COBE/FIRAS \cite{cobe}, i.e.~$\mu<6\times 10^{-5}$ at $95\%$ CL, and the constraint forecasted for an experiment like PIXIE \cite{Kogut:2011xw}, i.e.~$\mu<2 \times 10^{-8}$ at $95\%$ CL. More specifically, the plot shows $\mu$ as function of $\xic$, i.e.~the value of $\xi$ when the CMB pivot scale $k_{\ast}\equiv 0.002 \, {\rm Mpc}^{-1}$ left the horizon during inflation. In order to translate the constraints from TRIS plus COBE/FIRAS, we assume that the marginalized likelihood for $\mu$ is a Gaussian centered in zero with variance $\sigma_{\mu}^{2}=(3 \times 10^{-5})^{2}$, corresponding to the constraint $\mu<6\times 10^{-5}$ at 95$\%$ CL. Given that the model only gives positive $\mu$-distortion, we set the likelihood to zero for $\mu<0$. Using a flat prior on $\xic$, as discussed in section \ref{s:sum}, the $95\%$ CL constraint from TRIS plus COBE/FIRAS is $\xi_{c}<3.6$. Using a log-flat prior in the interval $\{10^{-1},10^{2}\}$, one finds $\xic<3.2$ at $95\%$ CL. 

An experiment like PIXIE, whose expected sensitivity is $\mu \simeq 2 \times 10^{-8}$ at $95\%$ CL, would be able to detect the $\mu$-distortion coming from this model. On the other hand, it would still not be possible to obtain a significant detection of non-vanishing $\xic$ from just this probe. The reason is that, even taking a fiducial signal as large as allowed by current data $\xic=2.41$ (see section \ref{s:an}), the amount of $\mu$-distortion is not significantly increased by the presence of the inverse decay effect as compared to the standard slow-roll case.

Summarizing, the constraint from $\mu$-distortion is currently weaker than those from the CMB spectrum and bispectrum to which we turn next.

%%%%%%%%%%%%%%%%%%%%%%%%%%%%%%%%%%%%%%%%%%%%%%%%%%%%%

\section{Bispectrum constraints}\label{ss:bi}

In this section we estimate the constraint on $\xic$ from the CMB temperature bispectrum. We follow the approach of \cite{BP} where the constraint from non-Gaussianity was first derived. We make a few different technical choices, which reflect in slightly different but comparable results. For the flat prior (see section \ref{s:sum}) on $\xic$ we find $\xic<2.45$ at $95\%$ CL, corresponding to $\fid<13$. For the log-flat prior (see section \ref{s:sum}) on $\xic$ we find $\xic<2.22$ corresponding to $\fid<0.4$ at $95\%$ CL. %EP Some technical details of the derivation are discussed in appendix \ref{a:bi}.

We gave the result for the primordial scalar bispectrum in section \ref{s:r} (details on the derivation are left to appendix \ref{a:bi}). By inspection the inverse decay bispectrum is largest in the equilateral limit, i.e.~when $k_{1}\sim k_{2} \sim k_{3}$. So instead of performing a dedicated bispectrum analysis we import and appropriately normalize the constraint on the equilateral non-Gaussian template. As first discussed in \cite{BP}, this can be done following \cite{Babich:2004gb}, i.e.~by computing the three-dimensional overlap of the two different shapes. This approach gives just an estimate of the non-Gaussian constraint on $\xic$, but given the similarity of the two shapes we expect it to be reasonably accurate. 

Let us start with a bispectrum
\be\label{defh}
\langle \R(\vec k_{1})\R(\vec k_{2})\R(\vec k_{3})\rangle=(2\pi)^{3} \delta \left(\sum_{i} \vec k_{i}\right) f_{NL} F( k_{1}, k_{2}, k_{3})\,.
\ee
To fix the ambiguity of moving numerical factors between $F$ and $f_{NL}$ in this definition we impose
\be\label{norm}
F(k_{\ast},k_{\ast},k_{\ast})=6\,\frac{3}{5} P_{\R}(k_{\ast})^{2} ,
\ee
where $k_{\ast}=0.002\,{\rm Mpc}^{-1}$ and the numerical factors are chosen to agree\footnote{One needs to take into account the conversion factor $\zeta=3 \Phi/5$.} with the way $f_{NL}$ is defined in numerical analysis, e.g.~\cite{Komatsu:2010fb,smith}. Before applying this to \eqref{eq:bispectrum} let us consider what the leading deviations from scale invariance are. $f_{3}$, discussed in appendix \ref{a:bi}, depends in principle on each one of $k_{1}$, $k_{2}$ and $k_{3}$ separately. There is also a dependence on $k_{i}$ in $e^{6\pi \xi}$ and $\Dsrki$, where the scale dependence of $\xi$ to first order in slow roll is given in \eqref{xifscale}. Let us estimate the dependence on an overall change in scale, affecting all $k_{i}$ in the same way. For various terms appearing  in the three-point function one finds
\be
\frac{\partial }{\partial \log k_{i}} \log f_{3}\simeq\frac{\dot \xic}{H}\frac{\partial}{\partial\xi} f_{3}&\sim& -4\eta_{\ast}\,,\label{f3es}\\
\frac{\partial }{\partial \log k_{i}} \log e^{6\pi \xi(k)}&=&3\pi \xic \eta_{\ast}\,,\\
\frac{\partial }{\partial \log k_{i}} \log \Delta_{\R,{\rm sr}}^{6}(k)&=&3(n_{s}-1)=-3(2\es+\eta_{\ast})\,,
\ee
where for the estimate in \eqref{f3es} we have used the large-$\xi$ limit $f_{3}\propto \xi^{-8}$ derived in \cite{BP}. In the regime of interest\footnote{It is true that in our analysis $\xic$ is varied all the way to zero, but all the effects of the inverse decay are exponentially suppressed for $\xic\lesssim2$, so in that regime the small corrections due to running are irrelevant.} $\xi\gtrsim 2$ and $\es\sim\eta_{\ast}\sim 10^{-2}$, the exponential generates the strongest running. The second strongest running comes from the power spectrum. For example, in the slow-roll quadratic model $\es=\eta_{V}=\eta_{\ast}/2$ and hence the scale dependence can be quickly estimated using $\eta_{\ast}\simeq 2\es-\eta_{V}\simeq\es$. Finally, $f_{3}$ induces the mildest running. One can therefore safely neglect the scale dependence in $f_{3}$ which can then be approximated by $f_{3}(\xic,k_{2}/k_{1},k_{3}/k_{1})$ defined in \eqref{f3x}. On the other hand, in the following we will keep the scale dependence in $e^{6\pi \xi}$ and $\Dsrki$. 

Using the normalization \eqref{norm} and \eqref{eq:bispectrum} one finds 
\be
\fid&=&\frac{\Delta_{\R,{\rm sr}}^{6}(k_{\ast})e^{6\pi\xic}f_{3}(\xic,1,1)}{\Delta_{\R}^{4}(k_{\ast})}\,,\\
F^{id}&=&\frac{3}{10}(2\pi)^{4}\Delta_{\R}^{4}(k_{\ast})\, \frac{\sum k_{i}^{3}}{\prod k_{i}^{3}}\,\frac{f_{3}(\xic,\frac{k_{2}}{k_{1}},\frac{k_{3}}{k_{1}})}{ f_{3}(\xic,1,1)}\left[\prod_{i}^{3} \left(\frac{k_{i}}{k_{\ast}}\right)^{\pi \xic \eta_{\ast}+n_{s}-1}\right]\,,
\ee
where we have collected all the deviation from scale invariance inside the shape function $F^{id}$. The equilateral template used by the WMAP collaboration is defined as
\be
F^{eq}&\equiv&\frac{3}{10}(2\pi)^{4}\Delta_{\R}^{4}(k_{\ast}) 3 \left[-\frac{1}{k_{1}^{3}k_{2}^{3}} \left(\frac{k_{1}k_{2}}{k_{\ast}^{2}}\right)^{n_{s}-1}-2\,{\rm perm's}+\right.\\
&&\quad \left. -\frac{2}{ k_{1}^{2}k_{2}^{2}k_{3}^{2}} \left(\frac{k_{1}k_{2}k_{3}}{k_{\ast}^{3}}\right)^{2(n_{s}-1)/3}+\frac{1}{k_{1}k_{2}^{2}k_{3}^{3}} \left(\frac{k_{1}k_{2}^{2}k_{3}^{3}}{k_{\ast}^{6}}\right)^{(n_{s}-1)/3}+ 5\,{\rm perm's}\right]\, \nonumber
\ee
and has been constrained to $-214<f^{eq}_{NL}<266$ at $95\%$ CL using WMAP 7-year data \cite{Komatsu:2010fb}.

A normalized scalar product, also known as ``cosine'', can be defined \cite{Babich:2004gb} between two shapes $F_{1}$ and $F_{2}$ by
\be
\cos \left(F_{1},F_{2}\right)\equiv \frac{F_{1}\cdot F_{2}}{\sqrt{F_{1}\cdot F_{1}} \sqrt{F_{2}\cdot F_{2}}}\,,
\ee
where the dot product is defined to mimic the scaling\footnote{Numerically one finds that introducing or not the weight factor $(k_{1}+k_{2}+k_{3})^{-1}$ changes the result less than $1\%$.} of the optimal three-point function estimator \cite{FS}
\be
F_{1}\cdot F_{2}\equiv\int_{k_{\rm min}}^{k_{\rm max}} dk_{1}d k_{2}dk_{3}\,\frac{\left(k_{1}k_{2}k_{3}\right)^{4}}{k_{1}+k_{2}+k_{3}}\,F_{1}(k_{1},k_{2},k_{3})\,F_{2}(k_{1},k_{2},k_{3})\,.
\ee
Here $k_{\rm max}$ and $k_{\rm min}$ are the smallest and largest scales relevant for a given experiment. The cosine can be used to import bounds from one shape, e.g.~the equilateral template, to another, e.g.~the inverse decay shape, which has not yet being compared with the data. After appropriately taking into account the potentially different ``volume'' of the two shapes, from a given constraint on $f_{NL}^{eq}$ one deduces a constraint on $f_{NL}^{id}$ using \cite{Babich:2004gb}
\be\label{fudge}
\Delta \fid = \frac{\Delta f_{NL}^{eq} }{\cos \left(F^{eq},F^{id}\right)} \sqrt{\frac{F^{eq}\cdot F^{eq}}{F^{id}\cdot F^{id}}}\,.
\ee
Numerically one finds $\cos \left(F^{eq},F^{id}\right)=0.935$ and $F^{eq}\cdot F^{eq}/F^{id}\cdot F^{id}\sim 5$. While the result for the cosine is very robust, the ratio of the volumes depends mildly on $k_{\rm max}$. As we vary it e.g.~from to $0.03$ to $0.05 \,{\rm Mpc}^{-1}$ (corresponding to $l\simeq400$ and $l\simeq700$, respectively), the volume ratio changes from $4.3$ to $5.3$. For the following estimates we take $F^{eq}\cdot F^{eq}/F^{id}\cdot F^{id}= 5.3$ corresponding to the largest range of scale relevant for WMAP, which results in the most stringent bound we could possibly set. Even with this conservative estimate, as we will see, the constraint is still weaker than the one from the power spectrum. 

Using these results, the $95\%$ CL constraint on the inverse decay template is $-100<\fid<124$. Notice that we have rescaled both the error bars and the central value by the fudge factor in \eqref{fudge}, so that the statistical significance of the Gaussian null hypothesis is not changed\footnote{This is important since otherwise the constraint on $\xic$ would depend on  the normalization of $F^{\rm id}$, which is unphysical.}. Despite the fact that the constraint on $\fid$ is weakened with respect to that on $f^{eq}_{NL}$ by the cosine being smaller than one, the inverse decay shape has five times more ``volume'', resulting in an overall \textit{stronger} bound than that for the equilateral template. Let us emphasize that the bounds quoted above assume a flat prior on $f_{NL}$. This is a very reasonable prior when looking in the data for any deviation from the standard picture. On the other hand, in this work we are considering a specific class of models, axion inflation, and hence we have theoretically motivated priors which are quite different (see section \ref{s:sum}). 

\begin{figure}
\centering
\includegraphics[width=.6\textwidth]{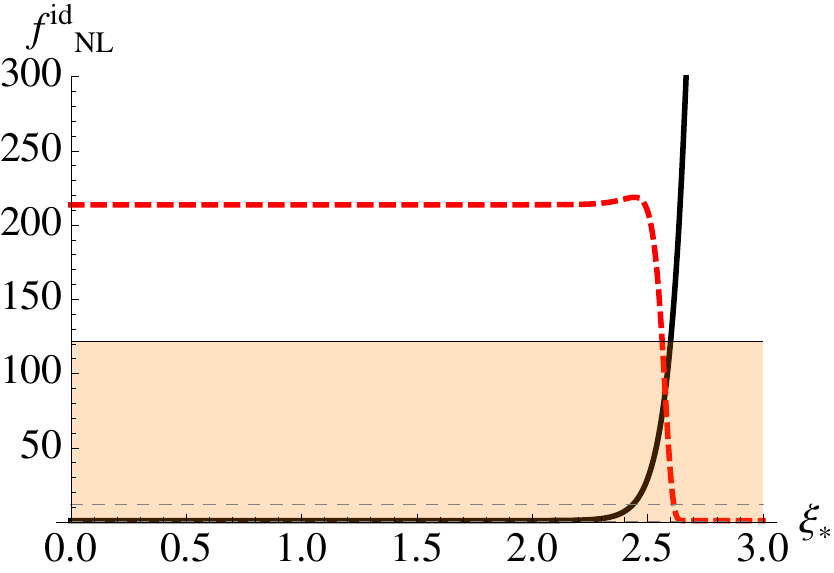}
\caption{As function of $\xic$, the plot shows $\fid$ (continuous black line) and the likelihood produced by the constraints on non-Gaussianity (dashed red line). For $\xic<2.4$, $\fid\sim0$ and the likelihood is completely flat. For $\xic\gtrsim 2.7$, $\fid\gg100$ and the likelihood drops to zero. The small peak corresponds to $\fid=12$ (thin dashed line), which is the central value for equilateral non-Gaussianity rescaled by the fudge factor in \eqref{fudge}.\label{ffnl}}
\end{figure}

To a good approximation the likelihood of $\fid$ is a Gaussian with standard deviation $\sigma_{\fid}=56$ and centered at $\fid=12$. In order to derive a bound on the parameter $\xic$ using $\fid(\xic)$, we have to take into account the fact that in the model under consideration $\fid(\xic)$ is strictly positive. We do this by setting the likelihood to zero for $\fid<0$. For a flat prior on $\xic$, we integrate $\mathcal{L}(\fid(\xic))$ in $d\xic$ over the interval $\{0-10\}$ . The resulting bound at $95\%$ CL is $\xic<2.45$ corresponding to $\fid<13$. For the more theoretically motivated log-flat prior (discussed in \ref{s:sum}) we integrate $\mathcal{L}(\fid(\xic))$ in $d\log \xic$ over the interval $\{10^{-1},10^{2}\}$. We find $\xic<2.22$ corresponding to $\fid<0.4$ at $95\%$ CL. As we will see in the next section, these bounds are weaker than those obtained from the CMB temperature power spectrum.

The reader might wonder why the bounds on $\fid$ are much stronger assuming a flat and even more a log-flat prior on $\xic$ rather than a flat prior on $f_{NL}$ as done in e.g.~\cite{Komatsu:2010fb,smith}. The reason is clear. For most values in the space of fundamental parameters, say $\alpha$ and $f$ or equivalently $\xic$, the model gives either no non-Gaussianity or is completely incompatible with the data. Only in a narrow window around $\xic\simeq 2.5$ can one have a potentially observable signal that has not yet been ruled out. So, within this model, $\fid\sim100$ is rather unlikely compared to $\fid\sim0$ or $\fid\gtrsim 10^{3}$. 
%Second, one might ask how robust this analysis is. On this topic it is worth mentioning that $\fid(\xic)$ is a very fast changing function of $\xic$, as can be seen figure \ref{ffnl}. Even if the fudge factor in \eqref{fudge} for the conversions of $f_{NL}^{eq}$ into $\fid$ bounds had been larger by a factor of two, the bounds on $\xic$ would not change more than $1\%$, showing that these results are fairly robust.

%%%%%%%%%%%%%%%%%%%%%%%%%%%%%%%%%%%%%%%%%%%%%%%%%%%%%%%%

\section{Power spectrum constraints}\label{s:an}

In this section we constrain $\xi_{\ast}$ by comparing the theoretical predictions for the temperature power spectrum with CMB data from the Wilkinson Microwave Anisotropy Probe (WMAP) and the Atacama Cosmology Telescope (ACT). We considered both the quadratic and generic models introduced in section \ref{s:sum}. The results are summarized in tables \ref{tab:analysis1} and \ref{tab:analysis2} for each model, respectively. Our main finding is that the strongest constraint comes from WMAP + ACT power spectrum for both the quadratic and generic model and that the constraint is stronger than the constraint from non-Gaussianity.

%The latter model requires 2 additional parameters, which should generally lead to a depletion of the constraints. However, in the model dependent analysis we explicitly add a non-zero tensor contribution that results in some interesting effects. 
%The best model independent constraint is derived when using WMAP and ACT data simultaneously, resulting in $\xi_* \lesssim 2.14$ at $95\%$ C.L. with a logarithmic prior on $\xi_*$. This is stronger than the bound derived from equilateral Non-Gaussianities. We also analyzed the performance of future experiments (Planck, ACTpol) for the the constraints on $\xi_*$. 

%%%%%%%%%%%%%%%%%%%%%%%%%%%%%%%%%%%%%%%%%%%%%%%%%%%%%

\subsection{Analysis}\label{s:an}

We modified the publicly available version of COSMOMC \cite{Lewis:2002} and CAMB to run with the power spectra discussed in section \ref{s:sum} for the quadratic and generic model, respectively. When running the Markov Chain Monte Carlo (MCMC) we took a conservative Gelman and Rubin bound of $R-1<0.01$ \cite{GR1992}. The Gelman-Rubin diagnostic $R$ relies on parallel chains to test whether they all converge to the same posterior distribution by considering the variance of the parameters in each chain compared to the variance of the same parameters over all parallel chains.  Convergence is diagnosed once the chains have `forgotten' their initial values, and the output from all chains has become indistinguishable ($R-1=0$).  In particular we aim at a sufficient convergence of $\xi_*$. To test convergence, we run 4 independent chains. For the quadratic model we computed the power spectrum numerically for a sufficient number of $k$ and $f_2$ (as a function of $\xi_*$). We used a 2D spline interpolation inside CAMB to call for arbitrary combination of $k$ and $\xi_*$. Note that the approximation used to compute $f_2$ in \cite{BPP} is only valid for $\xi_*>1$. We used the exact Whittaker solution in \eqref{Whittaker} in order to extend the range to $\xi_*=0$. For analyzing the generic model we use \eqref{ps}, to first order in slow-roll. Again, we call $f_2$ as a function of $\xi_*$ through a 1D spline.

In principle some issues could arise when analyzing the generic model of section \ref{s:sum}. If $n_s = 1$ then $\epsilon_* = -\eta_*/2$, which allows for the solution $\eta_* = \epsilon_* = 0$. In this case the power spectrum is exactly scale invariant to first order in the slow-roll expansion. Hence the constraint on $\xi_*$ will be undetermined (the correction to the single-field slow-roll solution is simply a renormalization of the total power). For WMAP and ACT\footnote{In analyzing ACT data we also include the SZ amplitude $A_{SZ}$, the amplitude of clustered point sources $A_{C}$ and the amplitude of Poisson distributed point sources $A_{P}$.}, $n_s = 1$ is still allowed by the data, which would suggest we could run into this issue when scanning the multidimensional likelihood. In addition, we expect slower convergence due to a doubling of the initial parameters $\{\xic,\;A_s\}\rightarrow \{\xic, A_s, \epsilon_*, \; \eta_*\}$. Another potential issue becomes apparent only when considering lensing. For large enough $\{\eta_*,\xi_*\}$ the spectrum becomes extremely blue tilted at small angular scales. Obviously, such a combination is not allowed by current data, and in principle this combination should be rejected. However, when considering lensing, a too large power spectrum invalidates some of the approximations made (e.g. small deflection angles) in the lensing computation. Therefore, typically one sets an upper bound $\Delta_{\R}^{2}(k_{max})<10^{-7}$. In order not to exceed this bound, we would have to strictly limit our prior values of $\eta_*$ and $\xi_*$.  Unfortunately, we can not constrain the priors on $\eta_*$ and $\xi_*$ too much, since a large value of one of these two parameters alone is still allowed by the data. Hence we set a hard bound on the power spectrum of $\Delta_{\R}^{2}(k_{max})<10^{-7}$, i.e.~if the power spectrum becomes larger than this, we set it to be  $10^{-7}$. 

The quadratic model generically predicts tensor modes, which we include in the data analysis. We assume the slow-roll result $\Delta^{2}_h(k)= 16 \epsilon_* \Dsr(k/k_*)^{n_t}$ and $n_t = -2\epsilon_*$. For the generic model, the constraints on the contribution from tensor modes to the temperature power spectrum, allows to break the degeneracy between $\epsilon_*$ and $\eta_*$.  The prior on $\eta_*$ is set such that $\xi>0$ for all observable $k$.  Let us now discuss the priors on $\xic$ in some more detail.

%%%%%%%%%%%%%%%%%%%%%%%%%%%%%%%%%%%%%%%%%%%%%%%%%%%%%

\subsection{Priors}

Before we discuss our results, let us briefly remind ourselves of the concept of bayesian analysis. In bayesian terminology we are interested in extracting the values of our parameters given the data, i.e.
\begin{eqnarray}
P(\Theta_i | D) & = &\frac{P(D|M(\Theta_i) )P(\Theta_i)}{P(M)}. 
\end{eqnarray}
Here $D$ represent the data, $\Theta_i$ are a set of parameters describing a model $M$. $P(\Theta_i)$ describes the parameter priors , while $P(D|M(\Theta_i))$ is known as the Likelihood function $\mathcal{L}$, i.e.~the probability of the measured data given the model $M$ with parameters $\Theta_i$. For the CMB this translates to \cite{Verde:2003ey}
\begin{eqnarray}
P(\Theta_i |  \hat{C}_l) & = & \mathcal{L}(\hat{C}_l  |  C_{l}^{\mathrm{th}}(\Theta_i)) P(\Theta_i),
\label{eq:probcl}
\end{eqnarray} 
where $\hat{C}_l$ is the best estimator of the true $C_l$, and $C_l^{\mathrm{th}}$ is the model predicted value of the power spectrum, given the parameters $\Theta_i$. The denominator $P(M)$ is of no interest as it does not depend on the parameters, in fact it equals the integral of the numerator over the parameter space. Consequently, it only affects the overall normalization of the posterior. 

One peculiarity of bayesian analysis is the choice of priors. Priors can be based on previous or independent experiments. For $\xi_*$ no such independent constraint exist, but we do have some theoretical prejudices on the value of $\xi_*$. It is usually a save bet to put a flat prior, i.e.~equal intervals in parameter space have equal likelihood, but it is most certainly not always a natural choice. In this case $\xi_*$ is not the parameter appearing in the fundamental Lagrangian describing the theory. The microscopic parameters are instead the axion decay constant $f$ and $\alpha$, which are related to $\xi_*$ through 
\begin{eqnarray}
\xi \equiv \frac{\alpha \dot{\phi}}{2fH}.
\end{eqnarray}
Within effective field theory we expect $\alpha = \mathcal{O}(1)$. Since $f$ is an energy scale, its value could range several orders of magnitude. For this reason, it would make more sense to expect log spaced intervals to be equally likely, i.e.~we should consider a log prior on $\xi_*$. Fortunately, as is apparent from eq.~\eqref{eq:probcl} it is straightforward to change priors after the computations of the likelihood through an MCMC (which is the computational intensive part). For completeness we consider both a flat and a flat log prior on $\xi_*$ and discuss the difference between the results in the next section.

%%%%%%%%%%%%%%%%%%%%%%%%%%%%%%%%%%%%%%%%%%%%%%%%%%

\subsection{Results}\label{ss:res}

\begin{figure}
  \centering
  \includegraphics[width=0.83\textwidth]{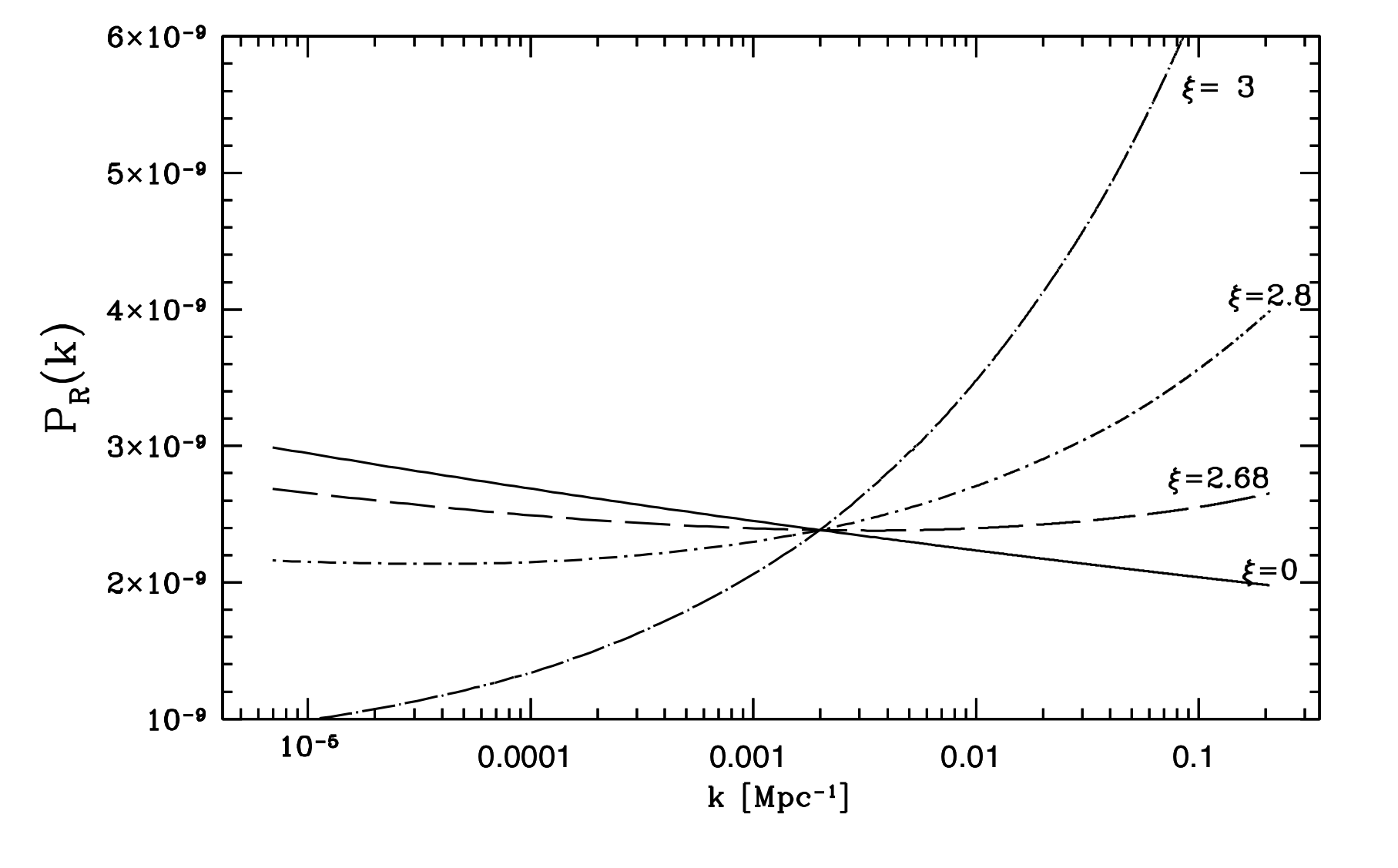}
  \caption{The primordial power spectrum for $\xi_*=0,2.68,2.8,3$ for a quadratic potential.}%EP
  \label{fig:prim_spectra}
\end{figure}

\begin{figure}
  \centering
\includegraphics[width=0.81\textwidth]{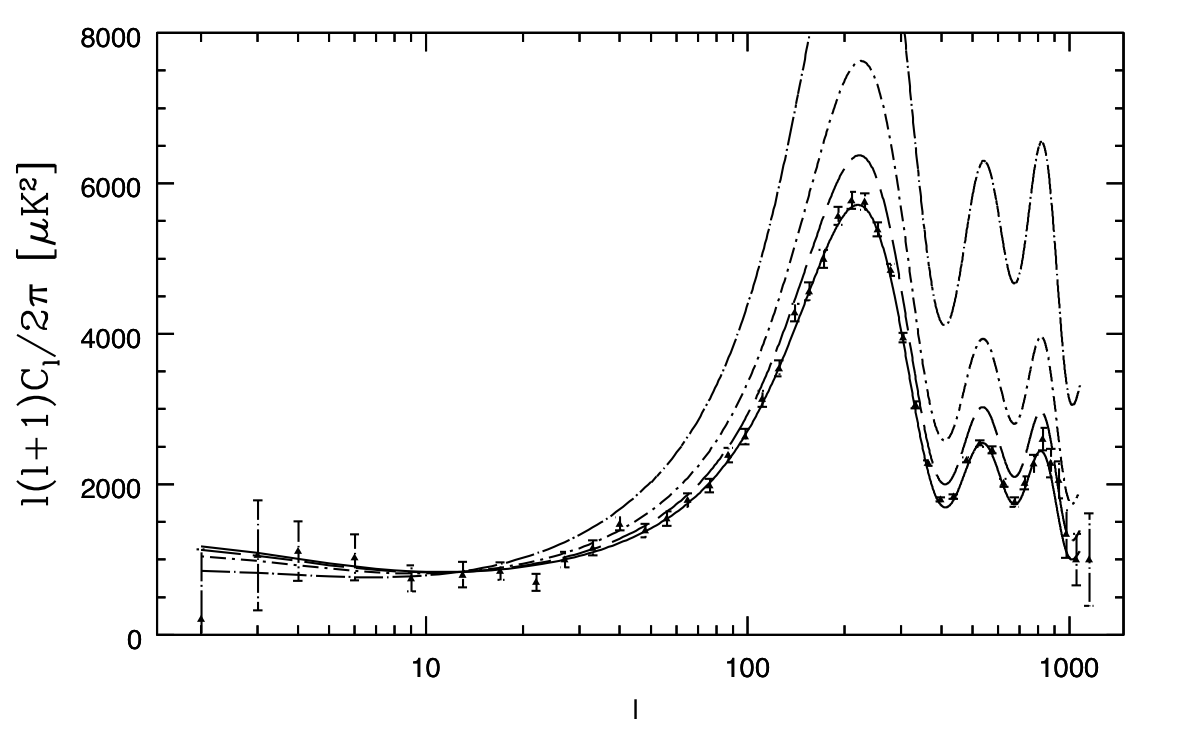}
  \caption{The late time spectrum for $\xi_*=0,2.68,2.8,3$ and using for all the other parameters the best fit values of $\Lambda$CDM (obtained for $\xic=0$). We also show the binned data points and errors (systematic and cosmic variance).} %EP Values of $\xi>2.6$ are excluded by current observations using best fit values for al other parameters. }
 \label{fig:late_spectra}
\end{figure}

In figures \ref{fig:prim_spectra} and \ref{fig:late_spectra} we have plotted both the primordial and late time power spectra for several values of $\xi_*$ and best-fit WMAP7 values of all other $\Lambda$CDM parameters. Large $\xi_*$ results in a poor fit for WMAP data. Because of the blue tilt generated by $\xic\gtrsim 2.5$, data on the power on small scales should further constrain the value of $\xic$. These small CMB scales have been probed by ground based experiments such as the Atacama Cosmology and South Pole telescopes (ACT \& SPT). We use ACT data and  our results are summarized in table \ref{tab:analysis1} (section \ref{s:sum}). We consistently account for the tensor contribution to the temperature power spectrum. From current CMB data we know that the best-fit model prefers $r=0$, so we expect to find a worsened likelihood compared to $\Lambda$CDM + $r$ for the quadratic model.

\begin{figure}
  \centering
  \includegraphics[width=0.8\textwidth]{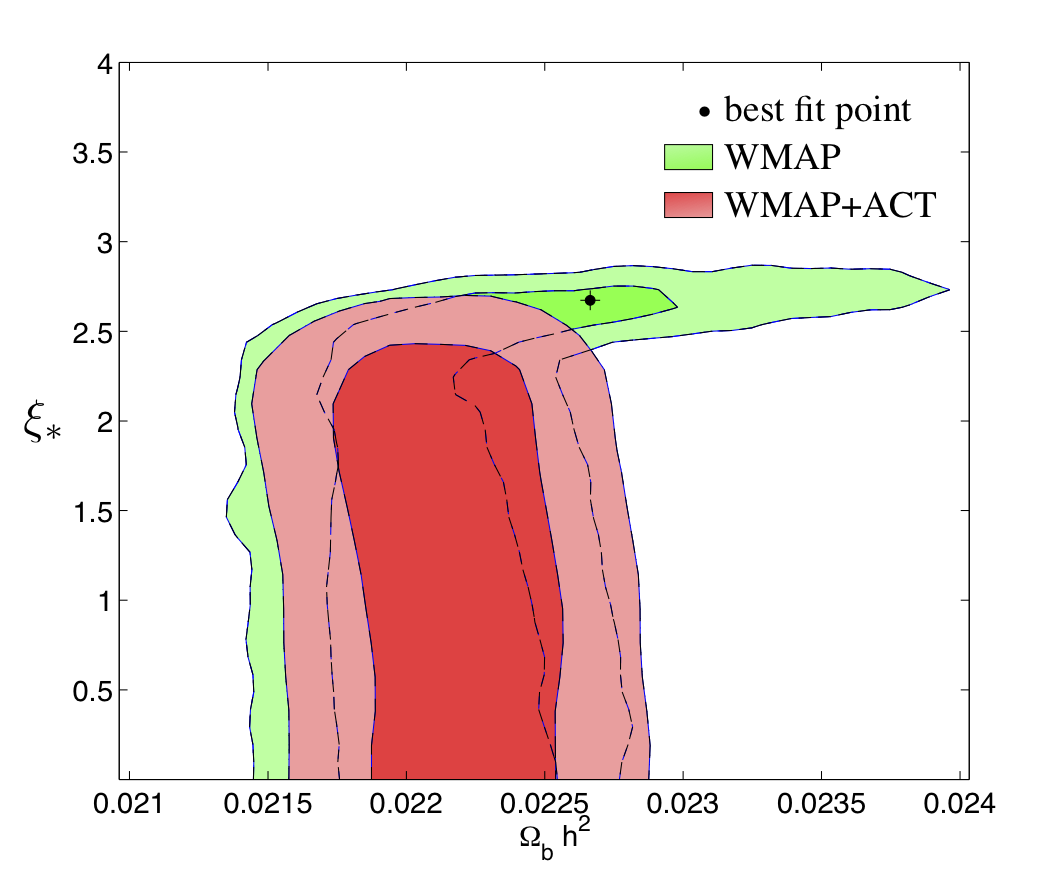}
  \caption{The two dimensional exclusion contours from WMAP and WMAP+ACT for $\Omega_{b} h^2$ versus $\xic$. WMAP alone allows for a better fit when both $\xic$ and $\Omega_b h^2$ are relatively large. This explains the weaker constraint.}%EP In fact, WMAP + tensors {\it prefers} a large value of $\xi$.}
  \label{fig:betsfit}
\end{figure}

\begin{figure}
  \centering
\includegraphics[width=0.81\textwidth]{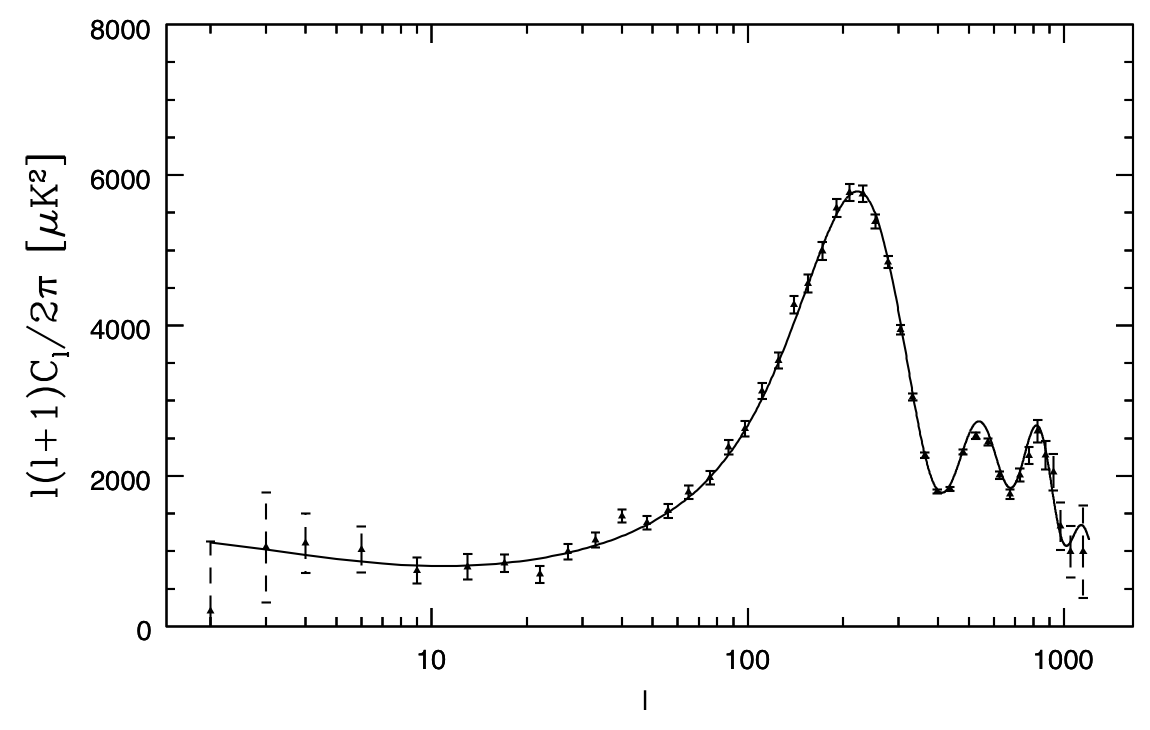}
  \caption{The best fit $C_l$ from the MCMC using WMAP data alone for a quadratic potential. Despite the relatively larger value of $\xi$ of 2.68, the increased value of $\Omega_b h^2$ and smaller $A_s$ can counteract this effect at small angular scales. Adding ACT, with a better hold on small scale BAO peaks, or putting the tensor contribution to zero will no longer prefer this weighted cancellation. }
 \label{fig:bestfitcl}
\end{figure}

\paragraph{Quadratic Model:} Combining  WMAP and ACT data yields the strongest constraint, i.e.~$\xic<2.41$ and $\xic<2.21$ for the flat and log-flat priors, respectively. Surprisingly, the constraint from WMAP alone is worse than naively expected. After inspecting the likelihood, we found that WMAP alone allows for a larger value of $\xi_*$. In fact, the best-fit point prefers $\xi_*=2.68$. First of all, the actual CMB data prefers no tensor modes, with a best fit WMAP7 value of $r=0$. For the quadratic model we include a non-zero $r$, which is  consistent with the models prediction. The tensor modes contribute to the low $l$, predominantly up to $l\sim 200$. To compensate for the increased power at low $l$ we can fit a smaller value of $\Delta_{\R,{\rm sr}}(k_{\ast})^{2}$, the primordial power. %EP
%Because the first peak is (much) more sensitive to tensor modes then the second and the third peak, to fit the first peak, 
The power is reduced mostly to fit the first peak, because that is where the WMAP data is most constraining. Now, the third peak and the second peak have too much power. To solve this $\Omega_b h^2$ is decreased, which leads to a shorter diffusion length, which equates into more damping on small scales. This is true in case $\xi_*$ is small. For larger $\xi_*$, the overall power can be reduced to fit the first peak and have relatively less effect on the second peak.  Although this might sound counter intuitive, note that the quadratic model has a red tilt, which causes the overall power enhancement in both peaks to be similar. The second peak and the third peak still have a little too much  power (as can be clearly seen from figure \ref{fig:bestfitcl}),  but fitting the second peak has less priority compared to the first, while the third peak is simply not measured well enough by WMAP. An improved measurement of the third to seventh peak with ACT reject this fitting, hence the relatively large improvement of the constraint on $\xi_*$. Figure \ref{fig:betsfit} shows a clear correlation between $\xi_*$ and $\Omega_b h^2$ for large values of $\xi_*$. A slightly smaller value of $\Delta_{\R,{\rm sr}}(k_{\ast})^{2}$, prefers a larger value of $\Omega_b h^2$ and $\xi_*$. In figure \ref{fig:bestfitcl} we show that the maximum likelihood point $\xic=2.68$ is in fact a reasonable good fit to the WMAP data. We also found that this correlation practically disappears without primordial tensors, which suggests that the improvement becomes redundant compared to an improvement from having zero tensor modes. We would like to stress that there are more factors involved in this best fitting than we have currently addressed. For example, there is also a role of $\Omega_{DM}h^2$, which seems to be correlated with $\xi_*$ as well. 

\begin{figure}
  \centering
  \includegraphics[width=0.8\textwidth]{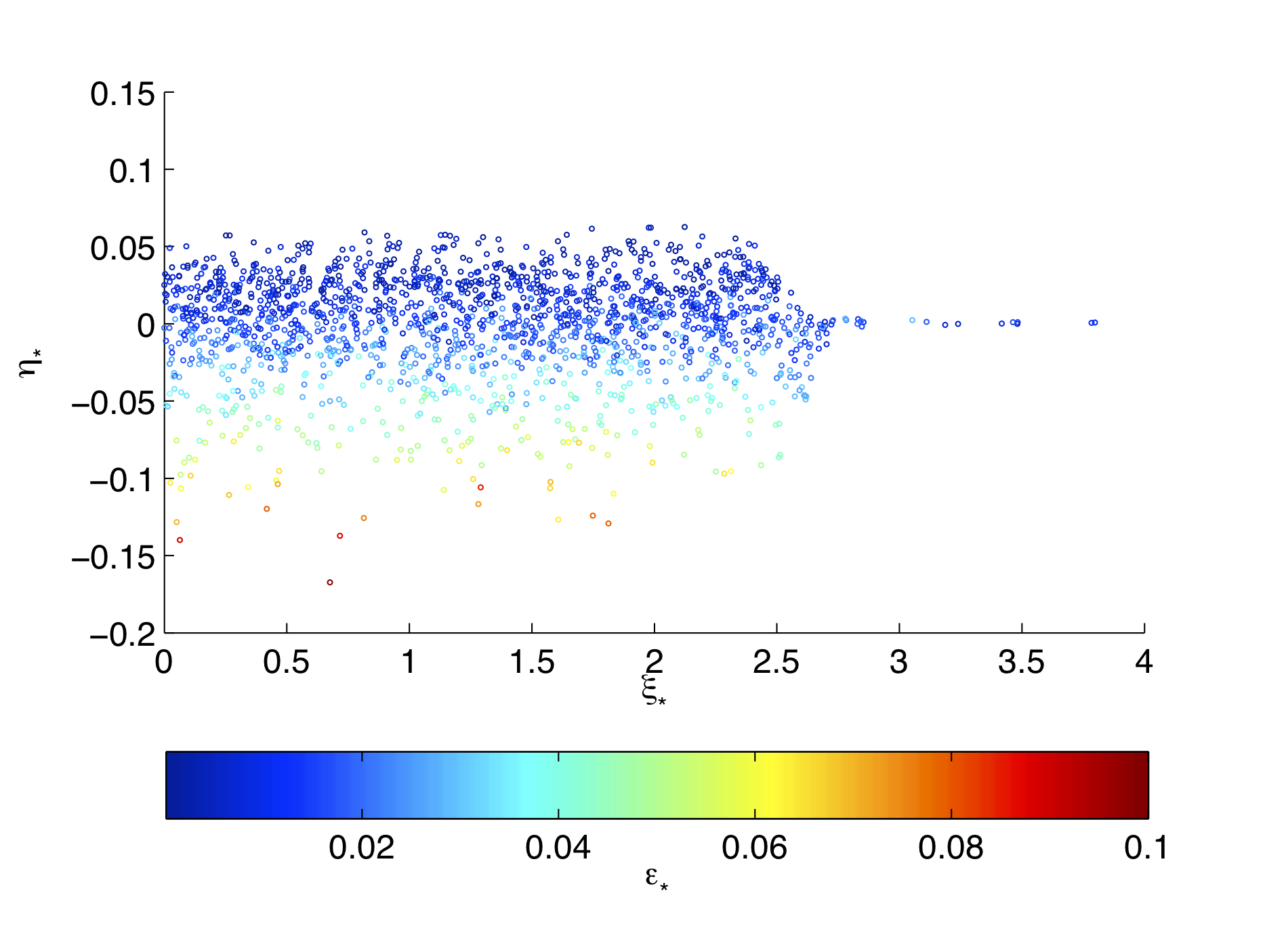}
  \caption{A multidimensional sample plot, showing `residual' samples around the line $\eta_* = \epsilon_* = 0$. }
  \label{fig:3dsample}
\end{figure}

\begin{figure}
  \centering
  \includegraphics[width=0.8\textwidth]{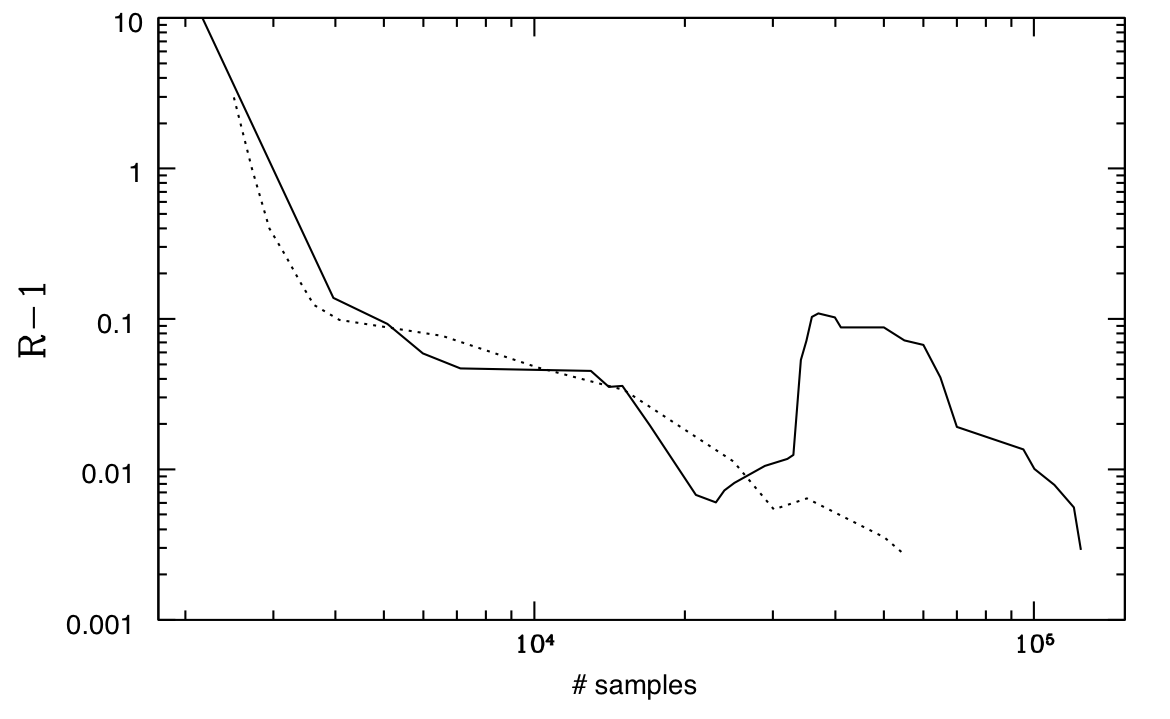}
  \caption{The Gelman-Rubin diagnostic as a function of the total number of samples for chains with a flat prior on $\xi_*$ (solid) and a flat log prior on $\xi_*$ (dashed) for the generic model. Chains with a flat prior have a tendency to `deconverge' once the chain is in the vicinity of $\eta_* = \epsilon_* = 0$. This results in varying constraints on $\xi_*$ from run to run. }
  \label{fig:GBbound}
\end{figure}

\paragraph{Generic Model: } As explained in section \ref{s:sum} for a generic model we introduce two extra parameters, $\{\epsilon_*,\eta_*\}$. Using WMAP plus ACT data, we find $\xic< 2.5$ and $\xic<2.14$ at $95\%$ CL for the flat and log-flat priors respectively. 
%EPBesides the addition of 2 parameters, the increase of the $2\sigma$ constraint for the flat prior is caused by sampling points around $\epsilon_{\ast} = \eta_* = 0$, where the inverse decay correction to the power spectrum becomes completely degenerate with the overall amplitude. 
Note that the constraint from WMAP is actually improved; the power in tensor modes is now a free parameter, and as explained the data prefers $r=0$ which removes the preference for large $\xi_*$. We did find that flat-prior constraints are relatively sensitive to the point $\epsilon_* = \eta_* =0$, in which case increasing $\xi_*$ just affects the overall amplitude and is therefore degenerate with $\Delta_{\R,{\rm sr}}(k_{\ast})^{2}$.  We qualify this in figure \ref{fig:3dsample}, where we show the 2 dimensional sample plot between $\eta_*$ and $\xi_*$ as a function of the value of $\epsilon_* $. There exist some residual points right around $\eta_* =\epsilon_* = 0$ precisely due to this effect.  The constraints are therefore relatively unstable. We ran multiple series of chains for which we found that the GR diagnostic was not satisfied despite the use of large samples because the chains tend to wonder around $\epsilon_* = \eta_*=0$ for extensive periods. Consequently, the chains gradually `de-converge'. This is shown in figure \ref{fig:GBbound}, where we plot the `route' to convergence as a function of the total number of samples. Eventually, the chains do converge, but the results differ slightly from {\it run to run}. Because of the this we do not trust the third digit in the constraints on $\xic$ with the flat prior. On the other hand, when applying the log prior to $\xic$ this effect is partly circumvented, since large values of $\xic$ are relatively unlikely. Instead of applying the log prior in postprocessing the chains, we sampled $\xi_*$ with a log-flat prior while running the MCMC. %EP Applying the log prior in the MCMC can also miss a signal when the signal is very local in likelihood space. In principle, running the chains for sufficient long times should recover the underlying distribution (the true posterior should be insensitive to the prior). However, even with a very conservative G-R bound, we found that it can be hard to recover the true posterior when looking for a non-zero $\xi_*$ in the data. This is explained in the next section, where we consider a fiducial signal in generated mock $C_l$.

After inspecting the chains with a log-flat prior on $\xi_*$ we found that, on average, about 1 chain out of 4 spoiled smooth convergence, resulting in long periods of no improvement in convergence as opposed to deconvergence for the flat prior. For our final constraints we only considered the chains that show consistent values for $\xi_*$ as a function of the percentage of samples we ignored, i.e.~changing the fraction from 20$\%$ to 50$\%$ changed the final constraint $\xi_*$ less then a few percent. %EP

Summarizing, the robust bounds are obtained using a log-flat prior in the MCMC and result in $\xic<2.18$ and $\xic<2.14$ for WMAP and WMAP plus ACT respectively at $95\%$ CL.

%%%%%%%%%%%%%%%%%%%%%%%%%%%%%%%%%%%%%%%%%%%%%%%%%%%%

\section{Future power spectrum constraints}

Planck data as well as small scale data from ACTpol will be soon available. In this section we forecast the constraints on $\xi_*$ that this new data will allow. The main results are summarized in table \ref{tab:analysis3}.

%%%%%%%%%%%%%%%%%%%%%%%%%%%%%%%%%%%%%%%%%%%%%%%%%%%%

\subsection{Analysis}

In order to construct a mock datasets we modified an existing code by Laurance Perotto and Julien Lesgourgues \cite{Perotto:2006rj}, which simulates the noise and cosmic variance from spherical harmonic $a_{lm}$, randomly generated from a fiducial signal plus noise spectrum distribution $\tilde{\mathcal{C}}^{ij}_l \equiv C_l^{ij}+N_l^{ij}$ , i.e.
\begin{eqnarray}
a^T_{lm} &=& \sqrt{ \tilde{\mathcal{C}}^{TT}_l} G^{(1)}_{lm}\,,\\
a^E_{lm} &=& \frac{ \tilde{\mathcal{C}}^{TE}_l}{ \tilde{\mathcal{C}}^{TT}_l} 
\sqrt{ \tilde{\mathcal{C}}^{TT}_l} G^{(1)}_{lm}
+\sqrt{\tilde{\mathcal{C}}^{EE}_l-\frac{(\tilde{\mathcal{C}}^{TE}_l)^2}{\tilde{\mathcal{C}}^{TT}_l}}
G^{(2)}_{lm}\,,\\
a^d_{lm} &=& \frac{ \tilde{\mathcal{C}}^{Td}_l}{ \tilde{\mathcal{C}}^{TT}_l} 
\sqrt{ \tilde{\mathcal{C}}^{TT}_l} G^{(1)}_{lm}+\sqrt{\tilde{\mathcal{C}}^{dd}_l-\frac{(\tilde{\mathcal{C}}^{Td}_l)^2}{\tilde{\mathcal{C}}^{TT}_l}}G^{(3)}_{lm}\,.
\label{alms}
\end{eqnarray}
Here $G^{k}$ are Gaussian random numbers with variance 1 and $T$= temperature, $E$ = $E$-mode polarization and $d$ is deflection due to lensing. We do not consider $B$-mode polarization.

One note of interest.  We found that {\it not} implementing noise through randomly generated spherical harmonic coefficients $a_{lm}$, seems to systematically underestimate the power when running an MCMC. This underestimation is particularly troublesome for our forecast in case of a non-zero $\xi_*$, which introduces a running that increases power on small angular scales. When simply adding white noise to the fiducial $C_l$, through a random Gaussian with variance 1, we found that the MCMC would consistently derive a lower value of $H_0$, driving $\xi_*$ away from its fiducial value. One explanation would be that we try to a fit a Gaussian likelihood to an actual likelihood that might not be Gaussian, which could lead to an underestimation of the true likelihood value. Given the large deviation from the fiducial input values, we did not trust the derived confidence levels. On the other hand, the method we use, prescribed through \ref{alms}, does recover all fiducial input parameters within $1\sigma$, indicating that it does not suffer from the same issue.

We have build a mock data sets for Planck and ACTpol.  We have considered $f^{planck}_{sky}=0.85$ at 3 different low frequency bands (70, 100 and 143 Ghz) and $f^{ACTpol}_{sky}=0.6$ at 148 Ghz. The later is the wide field survey planned for ACTpol, with a temperature noise variance $\Delta T = 21 \mu$K/arcmin per 1.4' arcmin beam and $\sqrt{2}$ larger for polarization. The estimated Planck temperature noise level for the respective frequency bands is 12.8 $\mu$K, 6.8  $\mu$K and 6  $\mu$K per arcmin beam ( 14', 10' and 7.1'). We also considered $E$-mode polarization, with noise levels of 18.2 $\mu$K,  11.4  $\mu$K and 6  $\mu$K, again per arcmin beam. 

We created two different mock data sets. First. we assume $\xi = \xi_* = 0$, $\epsilon_* = 0$ and $\eta_* = 0.04$, resulting in red tilt of $0.96$. A reason to choose $\epsilon_*= 0$, which implies no tensor contribution, is driven by trying to avoid $\eta_*=\epsilon_*=0$, where $\xic$ becomes completely degenerate with the power spectrum amplitude\footnote{Alternatively, one could have also chosen $\epsilon_* = 0.02$ and $\eta_*=0$, but we found that this choice does not show the stability of the first choice.}. Planck should be sensitive enough to be able to exclude $n_s=1$ for an input of $n_s=0.96$. Although the mock data set does not contain any tensor signal, in the MCMC we consistently consider the tensor contribution to the temperature power spectrum, which is determined by $\epsilon_*$ (or $r$). This breaks the degeneracy of $\es$ with $\eta_*$. 

Secondly, we consider a value of $\xi_*$ near the 2$\sigma$ level of WMAP + ACT, i.e.~$\xic\sim 2.4$.  Note that there is no need to generate $B$ modes since $B$ modes by themselves do not improve the constraint on $r$ and hence on $\epsilon_*$.   

%%%%%%%%%%%%%%%%%%%%%%%%%%%%%%%%%%%%%%%%%%%%%%%%%%

\subsection{Results}

We ran 4 MCMC chains with mock data sets with $\xi_*=0$ for Planck and Planck + ACTpol. The results are shown in table \ref{tab:analysis3}. We find that in the absence of a signal, Planck and Planck + ACTpol improve on current constraints\footnote{ If one does not fit for $A_{SZ}$, we found there exists some induced degeneracy between $\Omega_b h^2$ and $\xi_*$, which results in a somewhat better fitting near larger values of $\xi_*$, similar to the effect we found for the quadratic model fitting due to tensor modes. This effect weakens the constraint. In addition, we created a mock set with the same experimental characteristics of ACTpol, but with $l_{min} = 2$. We found that even though the beam is much smaller for ACTpol compared to Planck, the larger error and the use of a single band, actually lead to a weaker constraint on $\xi_*$ than Planck alone. }. %Given the unstable nature of the constraints using a flat prior in the generic model analysis (see discussion in section \ref{ss:res}), we prefer the constraints with log prior on $\xi_*$. Because in the mock data $r=0$, the results presented here are stable for the flat prior.

Adding a fiducial signal of $\xi_*=2.4$ to the mock datasets, is easily detected in both Planck and Planck + ACTpol analysis, with a very sharply peaked marginalized distribution. There is some skewness in the error bars due to the correlation with $\eta_*$. The addition of fiducial ACTpol data reduces the $2\sigma$ error bars by about $25\%$. This analysis shows that a signal sufficiently far away from zero could be detected by the Planck satellite and adding ACTpol will probably allow a detection of a weaker signal (closer to zero). Changing from flat to log prior does not change the constraints due to the sharp peak of the likelihood around $\xic\simeq2.4$. Running a chain with a log prior require some special caution because of the sharpness of this feature in the likelihood landscape; in an MCMC approach it can be easily missed, and one would incorrectly obtain some strong upper bound on $\xic$. Even with a very conservative G-R bound, one should allow for a sufficient running of the chains to assure this sharp feature in the likelihood is recovered. A good diagnostic for the log-flat prior having missed the signal is an anomalously weak upper bound on $\xic$. We conclude that it is advisable to run both a flat and a log prior and check for consistency. 

The analysis in this section supports the fact that current constraints are consistent with $\xi_*=0$, since a mock data set with 0 signal yields very similar results as the analysis of the real data. 

\begin{table}
\centering
\begin{tabular}{|c|c|c|c|} 
\hline\hline	
  & Planck & Planck + ACTpol &\tabularnewline
   \hline
flat prior (no signal) & $\xi_* < 2.17 $  &$\xi_* < 2.12 $  &\tabularnewline
 \hline
log prior (no signal)& $\xi_* < 1.96 $& $\xi_* < 1.92 $&	\tabularnewline
  \hline
flat prior ($\xi_* = 2.4$)& $ ^{+0.08}_{-0.07} $  & $^{+0.06}_{-0.04} $  &\tabularnewline
 \hline
log prior ($\xi_* = 2.4$)&$ ^{+0.08}_{-0.07}$ & $^{+0.06}_{-0.04}$&	\tabularnewline
 \hline 
 \end{tabular} 
 \caption{Constraints on $\xi_*$ at $95\%$ CL derived from Planck and ACTpol mock data with ($\xic=2.4$) and without ($\xic=0$) signal. For the mock data containing a signal we quote the $95\%$ CL error bar around the maximum likelihood point. }
 \label{tab:analysis3} 
 \end{table}

\section{Massive gauge field} \label{s:h}

It is interesting to ask what happens if the gauge field coupled to the inflaton gets a mass. In this section we consider a model in which this happens due to a Higgs-like field $h$ that develops a vacuum expectation value (vev). Perturbations of $h$ affect the efficiency of the tachyonic enhancement of $A$ which in turns changes the number of e-foldings. Therefore perturbations in $h$ are converted into curvature perturbations and can affect the late time observables. We point out a regime of parameters in which observably-large local non-Gaussianity can be generated remaining compatible with current data.

%%%%%%%%%%%%%%%%%%%%%%%%%%%%%%%%%%%%%%%%%%%%%%%%%%%%%%%%

\subsection{Equations of motion}

Let us start generalizing the action \eqref{L} with the addition of a complex scalar field $h$ charged under $A$ with some potential $W(h)$
\be\label{L2}
S=-\int d^{4}x \sqrt{-g}\left[\frac12 (\partial \phi)^{2}+|D h|^{2}+\frac14 F^{2}+\frac{\phi}{4f}F\tilde F+V(\phi)+W(h)\right]\,,
\ee
where $D_{\mu}\equiv\partial_{\mu}+ieA_{\mu}$. Notice that it is natural for the scalar potential to be additively separable because of the shift symmetry of $\phi$. For the effects we will consider, mixing with gravity gives just a small correction so we make the approximation of an unperturbed metric. This is tantamount to work in the spatially flat gauge, disregard vector and tensor modes and neglect the slow-roll suppressed interactions coming from the solution of the GR constraints. Taking the metric to be unperturbed FLRW, the six equations of motion plus one constraint are
\be 
\frac{1}{\sqrt{-g}}\partial_{\mu} \left[\sqrt{-g} \left(F^{\mu\nu}+\frac{\phi}{f}\tilde F^{\mu\nu}\right)\right]-2e^{2}A^{\nu}|h|^{2}+2e {\rm Im} \left(h\partial^{\nu} h^{\ast}\right)=0\\
\Box h-e^{2} A_{\mu} A^{\mu} h+i e \left[ 2A^{\mu}\partial_{\mu}h+ \frac{h}{\sqrt{-g}} \partial^{\mu} \left(A_{\mu} \sqrt{-g}\right)\right] + W_{,h}=0\\
\Box \phi - V'(\phi)-\frac{1}{4f} F_{\mu\nu}\tilde F^{\mu\nu} =0\label{eomphi}
\ee
with the usual definitions
\be 
\Box&\equiv& \frac{1}{\sqrt{-g}} \partial_{\mu} \sqrt{-g} \partial^{\mu}\,\\
\tilde F^{\mu\nu}&\equiv& \frac{\epsilon^{\mu\nu\rho\sigma}}{2 \sqrt{-g}}F_{\rho\sigma}\,,
\ee
with the totally antisymmetric tensor $\epsilon^{0123}=+1$. Focussing on the constraint, and choosing the Coulomb gauge $\nabla  \cdot \vec A=0$, one finds
\be\label{constraint}
a\partial_{i}^{2}A^{0}=\frac{1}{f} \nabla \phi \cdot \nabla \times \vec A + 2e^{2} A^{0}|h|^{2}+2e {\rm Im} \left(h\dot h^{\ast} \right)
\ee
We now consider the case in which $W(h)$ has a minimum for some $|h|={\rm const}$ and study the dynamics of the gauge field around the homogeneous background $\phi=\phi(t)$ and $|h|={\rm const}$. 

The following derivation parallels the ones in \cite{BP}. The constraint \eqref{constraint} is solved by $A^{0}=0$. Then the equations for the spatial components become
\be\label{eomA}
\vec A''-\partial_{i}^{2} \vec A-\frac{\phi'}{f} \nabla \times \vec A+2e^{2}|h|^{2} a^{2}\vec A=0\,.
\ee
Let us quantize the gauge field as
\be
\vec A(x,t)=\sum_{r=+,-} \int \frac{d^{3}k}{(2\pi)^{3}} \left[a_{r}(\vec k) A_{r}(k,t) e^{i \vec k \cdot \vec x} \vec \epsilon_{r}(\vec k)+{\rm h. c.}\right]\,,
\ee
where the polarization tensor obeys $\vec k \cdot \vec \epsilon_{\pm}(\vec k)=0$, $\vec k \times \vec \epsilon_{\pm}(\vec k)=\mp i k \vec \epsilon_{\pm} (\vec k)$, $\vec \epsilon_{\pm}(-\vec k)=\vec \epsilon (\vec k )^{\ast}$ and $\vec{ \epsilon}_{r}^{\ast}\cdot \vec \epsilon_{r'}=\delta_{rr'}$. Also
\be
\left[a_{r}(\vec k),a_{r'}^{\dagger}(\vec k')\right]=(2\pi)^{3}\delta_{rr'}\delta^{3}\left(\vec k-\vec k'\right).
\ee
\eqref{eomA} can be rewritten as
\be\label{eom}
\left[\frac{\partial^{2}}{\partial \tau^{2}}+k^{2}\pm \frac{2k\xi}{\tau}+\frac{m_{A}^{2}}{H^{2}\tau^{2}}\right] A_{\pm}=0\,,
\ee
where we defined $m_{A}^{2}\equiv 2 e^{2} |h|^{2}$. Without loss of generality, we assume $\xi>0$. Consequently the only mode that can undergo a tachyonic enhancement is $A_{+}$. We will consider this mode only and we will drop the subscript from here on.

\begin{figure}
\centering
\includegraphics[width=.45\textwidth]{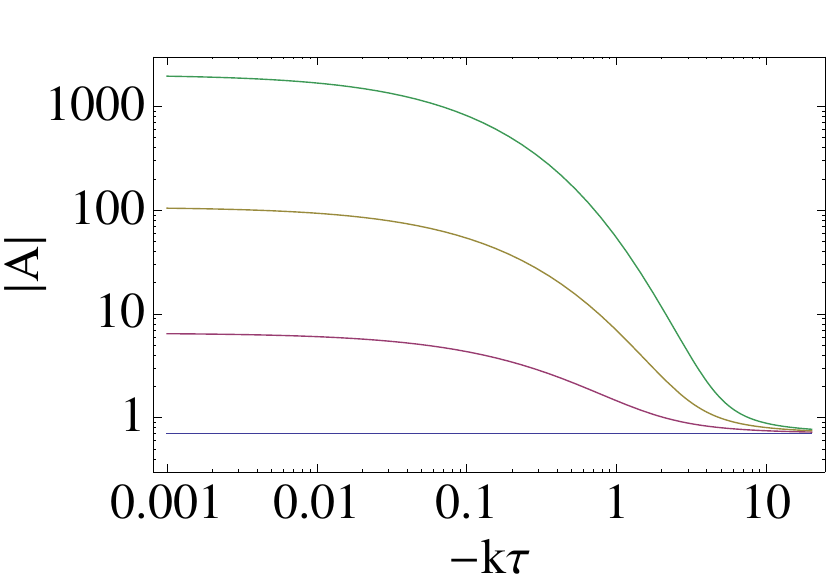}
\includegraphics[width=.45\textwidth]{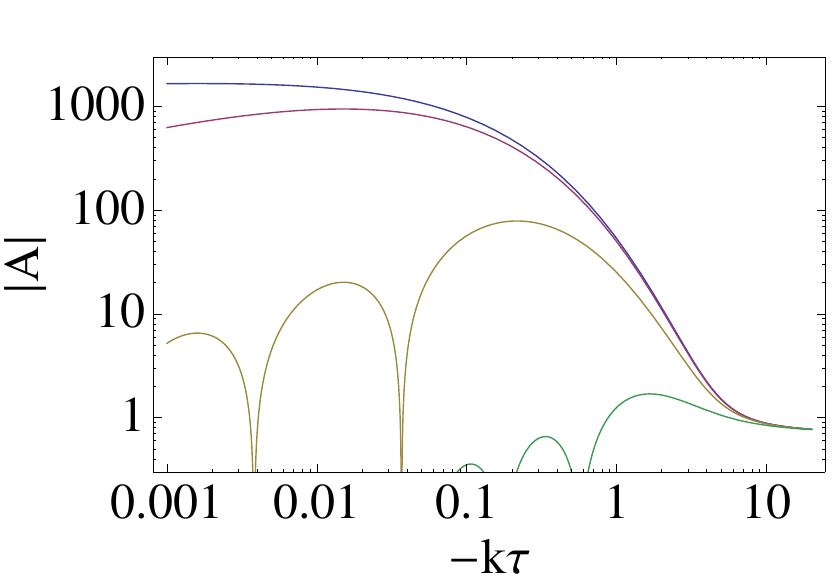}
\caption{The plots show $|A_{+}|$ in \eqref{A} as function of $-k\tau$ around horizon crossing $-k\tau=1$. In the left panel we have taken $m_{A}=0$ and $\xi=0,.2,.4,.6,.8,1$, in order to show the tachyonic enhancement. In the right panel we choose $\xi=3$ and $m_{A}/H=.2,.5,1.5,3$ in order to visualize how $m_{A}$ leads to a decay outside of the horizon and quenches the tachyonic enhancement already around horizon crossing. \label{figxi}}
\end{figure}

Before proceeding to find a formal solution of this equation it is useful to pause and understand the qualitative effect of the various parameters on the solution. Let us rewrite \eqref{eom} using $y=\log(-\tau)$
\be\label{int}
A_{,yy}-A_{,y}+ \left[(e^{y}k)^{2}-2(e^{y}k) \xi + \frac{m_{A}^{2}}{H^{2}}\right]A=0\,.
\ee
For $m_{A}=0$, which is the model studied in \cite{BP} and in the previous sections, $A$ undergoes an enhancement (exponential in $\xi$) around horizon crossing, before freezing out. This and the behaviors discussed in the following are confirmed in figure \ref{figxi}, which displays the numerical solution of \eqref{eom}. Intuitively this is analogous\footnote{Actually around and after horizon crossing, $e^{y}k = -\tau k \leq 1$, the frequency varies non-adiabatically, so this discussion should serve just to help intuition. From the exact solution presented in the following one can verify explicitly the qualitative behavior discussed here.} to a (damped) harmonic oscillator with an imaginary frequency $\omega^{2}\sim (e^{y}k)^{2}-2(e^{y}k) \xi$, i.e.~a tachyon. Far outside of the horizon $e^{y}k=-\tau k \ll1$, the friction term in \eqref{int} takes over and the solution asymptotes a constant. For small masses, i.e.~$m_{A}\ll H$, the exponential enhancement still takes place, but far outside of the horizon $A$ decays slowly (as an overdamped harmonic oscillator), $A\sim\tau^{\beta^{2}}$. As we increase $m_{A}$ for fixed $\xi$ two things happen. First, the tachyonic instability becomes smaller and smaller and eventually disappears when $m_{A}/H>\xi$. Second, the mode decays fast far outside of the horizon, $A\sim e^{y/2}k$.

A formal solution can be found by re-writing \eqref{eom} in the form of the Whittaker equation 
\be
\left[\frac{\partial^{2}}{\partial z^{2}}-\frac{1}{4}+\frac{\lambda}{z}+\frac{1/4-\mu^{2}}{z^{2}}\right]W_{\lambda,\mu}(z)=0\,,\label{Whittaker}
\ee
where $W_{\lambda,\mu}(z)$ is the Whittaker function. Using the large $|z|$ asymptotic behavior $W_{\lambda,\mu}(z)\rightarrow e^{-z/2} z^{\lambda}$ and the Bunch-Davies vacuum $A_{+}\simeq e^{-ik\tau}/\sqrt{2k}$, one can fix both integration constants in the general solution of \eqref{eom}
\be\label{A}
A_{+}=\frac{e^{\xi \pi/2}}{\sqrt{2k}} W_{-i\xi,\mu}(2ik\tau)\,,
\ee
where $\mu^{2}=1/4-m_{A}^{2}/H^{2}$. 

%%%%%%%%%%%%%%%%%%%%%%%%%%%%%%%%%%%%%%%%%%%%%%%%%%%%

\subsection{E-foldings}

In this subsection we want to estimate how the total number of e-foldings is affected by $m_{A}^{2}\equiv 2 e^{2} |h|^{2}$. This will allow us in the next subsection to estimate the conversion of $h$ into curvature perturbations using the $\delta N$ formalism. 

\begin{figure}
\centering
\includegraphics[width=.5\textwidth]{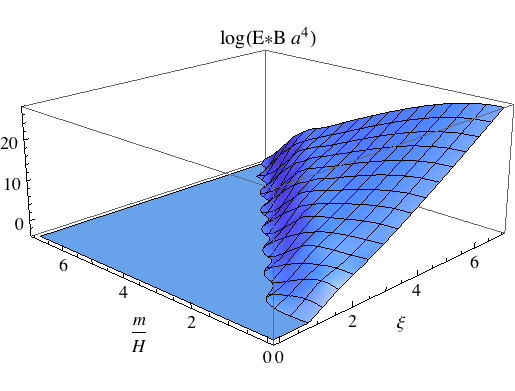}
\caption{The plot shows $\log \left(\ex{\edb} a^{4}\right)$ computed numerically as function of $m_{A}/H$ and $\xi$. A sizable contribution is generated only when $\xi>m_{A}/H$.\label{figedb}}
\end{figure}

The quantities that affect the total number of e-foldings $N$ are those appearing in the homogeneous equations \eqref{fh} and \eqref{Fried}: $-F\tilde F/4=\edb$ and $\vec E^{2}+\vec B^{2}$. They can be computed from
\be
\ex{\edb}&=&-\frac{1}{4\pi^{2} a^{4}}\int_{0}^{\infty} dk\,k^{3}\,\frac{\partial }{\partial \tau} |A_{+}|^{2}\,,\\
\frac{1}{2}\ex{\vec E^{2}+\vec B^{2}}&=&\frac{1}{4\pi^{2} a^{4}}\int_{0}^{\infty} dk\,k^{2}\, \left[|A'_{+}|^{2}+k^{2}|A_{+}|^{2}\right]\,,
\ee
where $\vec E\equiv - \vec A'/a^{2}$ and $\vec B\equiv \vec \nabla \times \vec A/a^{2}$. Using the variable $x\equiv-k\tau$, the solution \eqref{A} and the fact that the background is very close to de Sitter, which sets $a\simeq-1/(H\tau)$, we can rewrite these integrals as
\be\label{intedb}
\ex{\edb}&=&\frac{H^{4}}{8\pi^{2}}\,e^{\xi\pi}\int_{0}^{\infty} dx\,x^{3}\,\partial_{x} |W_{-i\xi,\mu}(-2ix)|^{2}\,,\\
\frac{1}{2}\ex{\vec E^{2}+\vec B^{2}}&=&\frac{H^{4}}{8\pi^{2}}\,e^{\xi\pi}\int_{0}^{\infty} dx\,x^{3}\, \left[|\partial_{x}W_{-i\xi,\mu}(-2ix)|^{2}+|W_{-i\xi,\mu}(-2ix)|^{2}\right]\,.
\ee
This form makes it clear that both quantities are slowly varying functions of time through $H(t)$ and $\xi(t)$. Both integrals are UV divergent, as could have been expected since we are multiplying fields at the same spacetime point. This UV divergence has nothing to do with the gauge field production, which takes place near the horizon exit of every mode. Following \cite{lorenzo,BP} we regularize these integrals by restricting the integration over the interval $0<x<2\xi$. One can check that $\ex{\edb}$ is always more important during inflation than $\ex{\vec E^{2}+\vec B^{2}}$ so that the latter can be neglected. In figure \ref{figedb} we plot $\ex{\edb}$ as function of $m_{A}/H$ and $\xi$. As previously argued, large tachyonic enhancement takes place only for $\xi>m_{A}/H$. 

\begin{figure}
\centering
\includegraphics[width=.4\textwidth]{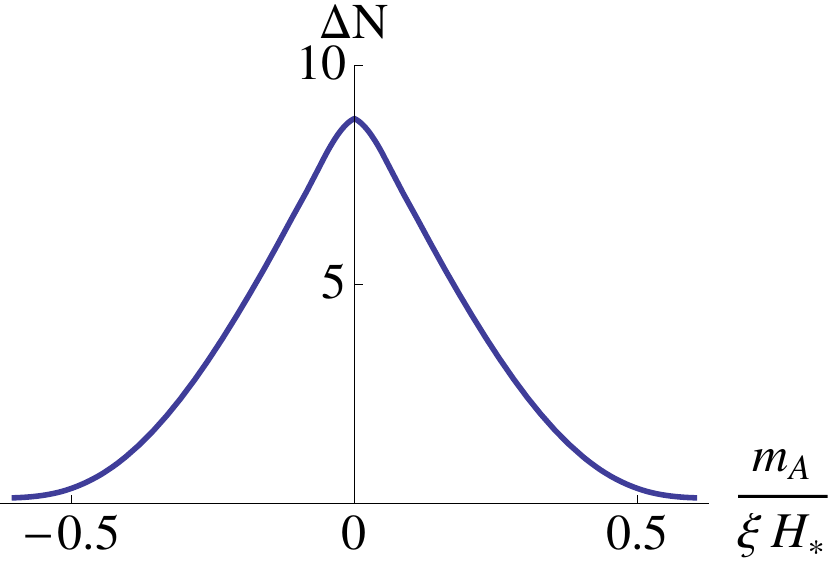} \qquad
\includegraphics[width=.4\textwidth]{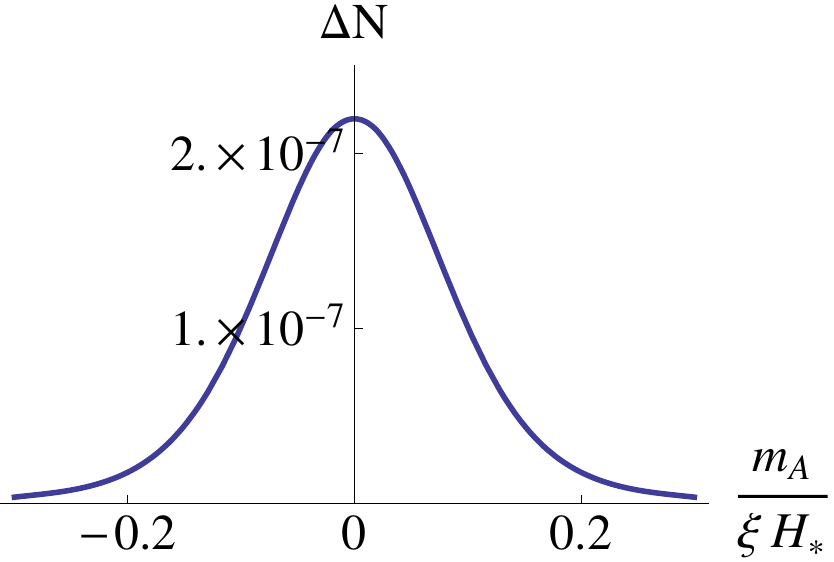}
\caption{The plots show the additional e-foldings $\Delta N$ due to strong-backreaction of $\ex{\edb}$ as function of the ratio $m_{A}/(\xi H_{\ast})$ for fixed $\xi=2.4$ on the left and $\xi=.3$ on the right. The plots are produced using a quadratic inflaton potential and $H_{\ast}=5 \times 10^{-5}$ is the Hubble parameter at some pivot time during inflation. This downward-bell shape can be easily understood. For $m_{A}=0$ the tachyonic enhancement is maximized and so is $\Delta N$. For $m_{A}/H \gg \xi$ the tachyonic enhancement is completely quenched and $\Delta N=0$. The plot is symmetric around zero since $\ex{\edb}$ is a function of $m_{A}^{2}$.\label{figN}}
\end{figure}

What we have learned so far allows us to understand how the total number of e-foldings $N$ depends on the parameter $m_{A}/H$. During inflation $\xi$ evolves slowly, being proportional to $\sqrt{2\epsilon}$. The regime we study in this paper is when $\xi(\tau)$ is small enough that $\ex{\edb}$ is a negligible correction to the background dynamics at the time cosmological perturbations $k<1\, {\rm Mpc^{-1}}$ left the horizon. By the end of inflation though, $\xi(\tau)$ can grow such that $\ex{\edb}$ starts backreacting on the homogeneous evolution by slowing down the inflation. If $N_{\ast}$ is the total number of e-foldings between the end of inflation and when some pivot cosmological scale $k_{\ast}$ left the horizon, we define
\be
\Delta N (\xi,m_{A}) &\equiv& N_{\ast}(\xi,m_{A})-N_{\ast}(\xi=0,m_{A}=0).
\ee
For example, in the case of a monomial inflation potential such as $m^{2}\phi^{2}$ and $m_{A}=0$ one finds $\Delta N(m_{A}=0) \sim \mathcal{O}(10)$ additional e-foldings due to this strong backreaction regime. On the other hand, as we can see from figure \ref{figedb}, if we increase $m_{A}/H$ from zero to $\xi$, the tachyonic enhancement of $\ex{\edb}$ disappears and the additional e-foldings with it, i.e.~$\Delta N(m_{A}/H \gtrsim \xi)\simeq 0$. Addiontionally we also know that $\ex{\edb}$ and therefore $\Delta N$ depend quadratically on $m_{A}$. Putting all of this together we deduce that $\Delta N(m_{A})$ must look like a downward bell, with maximum at $m_{A}=0$ and asymptotically approaching zero for $m_{A}/H\gg \xi$. This is confirmed by the numerical computation shown in figure \ref{figN}, where we plot $\Delta N$ for fixed $\xi$ as function of the ratio $m_{A}/(\xi H_{\ast})$.

%The only notable exception is when $\xi=m_{A}=0$, in which case $\ex{\edb}=0$. Using \eqref{A}
%\be 
%\lim_{k\rightarrow +\infty}k^{3}\,\frac{\partial }{\partial \tau} |A_{+}|^{2}=\frac{\xi k}{2\tau^{2}}-\frac{3\xi^{2}-\beta^{2}}{2\tau^{3}}+\dots
%\ee

%%%%%%%%%%%%%%%%%%%%%%%%%%%%%%%%%%%%%%%%%%%%%%%%

\subsection{Local non-Gaussianity}

In this subsection we use the $\delta N$ formalism \cite{dN,Lyth:2005fi} to estimate the conversion of perturbations of the Higgs field $h$ into curvature perturbations. We find that there is a regime of parameters in which observable local non-Gaussianity can arise. This regime requires the Higgs field $h$ to have a small mass compared to $H$, which technically, at the level of the present construction, is not a natural assumption.

The $\delta N$ formalism \cite{dN,Lyth:2005fi} provides a relation between curvature perturbations $\R$ and the different amount of expansion that different regions of the universe experience. More specifically, at scales much larger than the horizon  when gradient terms are negligible (separate universes approximation), curvature perturbations $\R$ evaluated at some time $t$ on a constant energy-density hyper-surface are equal to the perturbations in the number of e-foldings $\delta N$ between that same hypersurface and an initial flat hypersurface at $t_{i}$. When the number of e-foldings is a function of the value of a set of fields at time $t$, it can be expanded in a Taylor series. One finds
\be\label{dn}
\R(t)\simeq\sum_{I}N_{,I}\delta \varphi^{I}(t_{i})+\frac{1}{2}\sum_{IJ} N_{,IJ}\delta \varphi^{I}(t_{i}) \varphi^{J}(t_{i})+\mathcal{O} \left(\delta \varphi^{3}\right)\,,
\ee
where $N_{,I}\equiv \partial N/\partial \varphi^{I}(t_{i})$ and similarly for $N_{,IJ}$. In the model considered in this section \eqref{L2} we have two scalar fields $\phi$ and $h$ and the vector field $A$. When the backreaction of $\ex{\edb}$ is negligible at the time cosmological perturbations $k<1 {\rm Mpc^{-1}}$ leave the horizon, the perturbations in the gauge field do not affect curvature perturbations. This is the result of the fact that the integral in \eqref{intedb}, which determines the size of $\ex{\edb}$, is almost entirely supported around horizon crossing $x\sim \mathcal{O}(1)$. Perturbations in $A$ of cosmological size $k< 1 {\rm Mpc^{-1}}$ affect $\ex{\edb}$ only when this is a negligible correction to the homogeneous equations of motion. Later on, closer to the end inflation, $\ex{\edb}$ starts to dominate the homogeneous dynamics, but the modes that cross the horizon at that moment correspond to scales tens of orders of magnitude smaller than those relevant for cosmology.

Things are different for the scalar fields if they are lighter than $H$. The inflaton must be light because the shift symmetry is only weakly broken by the slow-roll flat inflaton potential. The mass of $|h|$ around its minimum (or $W(h)$ in \eqref{L2}) on the other hand is a free parameter. Not surprisingly interesting phenomenology arises only when $m_{|h|}\ll H$. Henceforth we will only consider this regime. The mass of $h$ is not protected by a symmetry and in that sense a small mass is unnatural. Let us estimate the various derivatives appearing in \eqref{dn} starting with the inflaton. $N_{,\phi}\simeq(2\epsilon)^{-1/2}\sim \mathcal{O}(10)$, and we can neglect $N_{,\phi\phi}$ since single field slow-roll inflation produces negligible non-Gaussianity \cite{Maldacena:2002vr,Acquaviva:2002ud}. The cross derivative $N_{,h\phi}$ is smaller than $N_{,hh}$ and therefore we neglect it as well. Moving on to derivatives with respect to $h$, we use the results of the previous subsection and $\partial_{h}=\sqrt{2}e\partial_{m_{A}}$. For $m_{A}\lesssim H \xi$ one can approximate $\Delta N(m_{A})$ as a downward parabola. One hence find the order of magnitude estimate
\be
N_{,h}\simeq \sqrt{2}e\partial_{m_{A}}N\sim e \frac{\dnm}{\xi H} \frac{m_{A}}{\xi H}\,,
\ee
where $\dnm\equiv \Delta N (m_{A}=0)$ is the largest amount of extra e-foldings for fixed $\xi$ and fixed potential (see e.g.~figure \ref{figN}). For the second derivative, a similar estimate gives
\be
N_{,hh}\simeq 2 e^{2} \partial_{m_{A}}^{2}N\sim e^{2}\dnm \frac{1}{(\xi H)^{2}}\,.
\ee
Combining these results we find the order of magnitude estimate
\be
\R\sim \frac{\delta \phi}{\Mpl}(2\epsilon)^{-1/2}+\frac{\delta h}{H} e \frac{\dnm}{\xi } \left[ \frac{m_{A}}{\xi H}+ \frac{e}{\xi} \frac{\delta h}{ H} \right]
\ee
where for massless fields we can use the standard result 
\be
\ex{\delta \phi^{2}}=\ex{\delta h^{2}}\simeq (2\pi)^{3} \delta^{3}(\vec K) \left(\frac{H}{2\pi}\right)^{2}\,.
\ee
A useful formula for local non-Gaussianity in the $\delta N$ approach \cite{Lyth:2005fi} is given by
\be
\floc=\frac{5}{6}\frac{\sum_{IJ}N_{I}N_{J}N_{IJ}}{\left(\sum N_{I}^{2}\right)^{2}}\,.
\ee
Imposing the COBE normalization $\Delta_{\R}^{2}(k_{p})=2.4\times 10^{-9}$, one obtains the following estimate for the late time local non-Gaussianity produced by the the conversion of $\delta h$ into $\R$
\be
\floc\sim 10^{2} \left(\frac{\dnm^{3/4}e}{\xi 10^{-3}}\right)^{4}\, \left(\frac{m_{A}}{\xi H}\right)^{2}\,.
\ee
One has to check that the contribution of $\delta h$ to $\R$ does not exceed the COBE normalization, this is guaranteed if
\be
\frac{\floc}{10^{7}}\frac{\xi^{2}}{e^{2}\dnm}\left(1+\frac{e H}{2\pi m_{A}}\right)<1\,.
\ee
It is now easy to check that observable\footnote{WMAP7 \cite{Komatsu:2010fb} imposes the bound $-10<f_{NL}^{\rm loc}<74$ at $95\%$ CL, and it is expected Planck will be sensitive to $\Delta \floc \sim 5$.} local non-Gaussianity can be generated in this model, for example $\xi\sim\dnm\sim1$, $m_{A}/(\xi H)\sim 0.1$ and $e=2\times 10^{-3}$ gives $\floc\sim 40$. Here we would like to emphasize that $\floc$ can range from very large to very small number and the volume of parameter space in which $\floc$ is detectable by say Planck but not yet ruled out by WMAP is a small fraction of the total volume.

%%%%%%%%%%%%%%%%%%%%%%%%%%%%%%%%%%%%%%%%%%%%%%%%

\appendix

\section*{Acknowledgments}

We are thankful to Neil Barnaby, Christian Wagner, David Spergel, Renee Holzek, and Matias Zaldarriaga for useful discussions. E.P. is supported in part by the Department of Energy grant DE-FG02-91ER-40671. P.D.M is supported by the Netherlands Organisation for ScientiÞc Research (NWO), through a Rubicon fellowship.

\section{On the scale dependence of the power spectrum}\label{a:run}

Following \cite{BP}  we will now re-derive the inverse decay corrections to the power spectrum \eqref{ps} with particular attention to the slow-roll deviations from scale invariance. 

Let us start considering the simple case of single field inflation, without any gauge field. One can expand the action around a homogenous slow-roll background to quadratic order in the perturbations. One can fix a flat gauge in which\footnote{We neglect vector and tensor perturbations.} $g_{ij}=a^{2}\delta_{ij}$. After solving the constraints coming from the Einstein equations, e.g.~as in \cite{Maldacena:2002vr}, one finds the equation of motion (see e.g.~\cite{Easther:2010mr})
\be\label{eomt}
\left[\frac{\partial^{2}}{\partial \tau^{2}}-\nabla^{2}+2aH\frac{\partial}{\partial\tau}+a^{2} \left(V''+16\pi G\frac{\dot \phi}{H}V'+8\pi G\frac{\dot\phi^{2}}{H^{2}}\right)\right]\dep=0\,.
\ee
We have written out the Newton constant explicitly to show which terms arise from the interaction with gravity through the constraint equations. In order to keep all slow-roll corrections into account, in \eqref{eomt} we should use\footnote{Notice that in the left hand side of \eqref{att} both $H$ and $a$ are functions of $\tau$, as can be seen by taking the derivative of both sides with respect to $\tau$. Unless a $\ast$ is present, all quantities are functions of time.}
\be
aH&=&-\frac{1}{(1-\epsilon_{\ast})\tau}+\mathcal{O}(\epsilon^{2})\,.\label{att}
\ee 
We also find it useful to use the variable $x\equiv-k\tau$ and quantize according to\footnote{Note that we have different conventions with respect to \cite{BP} in both the factors of $2\pi$ and of $a(t)$.}
\be
\dep(x,\tau)=\int \frac{d^{3}k}{(2\pi)^{3}} \left(e^{i\vec k\cdot \vec x}\, \dep_{k}\, b_{\vec k}+{\rm h.c.}\right)\,,\\
\left[b_{\vec k},b^{\dagger}_{\vec k'}\right]=(2\pi)^{3} \delta \left(\vec k-\vec k'\right)\,.
\ee 
Then \eqref{eomt} can be rewritten as
\be\label{eomfx}
\left[\frac{\partial^{2}}{\partial x^{2}}+1-\frac{2 \left(1+\es\right)}{x}\frac{\partial}{\partial x}+a^{2}H^{2} \sr\right]\dep=0\,,
\ee
where $\sr=3 \left(3\es-\eta_{\ast}/2\right)$ is some linear combination of $\es$ and $\eta_{\ast}$, which as we will see is not relevant to compute deviations from scale invariance at leading order in the slow-roll parameters. A positive frequency solution of \eqref{eomfx} is proportional to $x^{3/2+\es} H^{(1)}_{\nu}(x)$ where $\nu=3/2+\mathcal{O}(\es,\eta_{\ast})$ and the exact expression for the slow-roll correction is again not important for our purposes. Imposing the Bunch-Davies vacuum at some initial moment $x=x_{i}\gg1$ one can fix the (second) integration constant $\cq$
\be
\dep_{k}=\cq \,x^{3/2+\es} H^{(1)}_{\nu}(x)\rightarrow  \cq x_{i}^{1+\es}  e^{i x_{i}} \sqrt{\frac{2}{\pi}}=\frac{ H(\tau_{i}) }{\sqrt{2k^{3}}}x_{i}e^{i x_{i}}
\ee
up to an irrelevant phase. Note here that $H^{(1)}$ is the Hankel function, not to be confused with the Hubble parameter $H$, appearing on the right. The time dependence of the Hubble paramater $H$ at linear order in the slow-roll parameters is
\be
H(\tau_{1})=H(\tau_{2}) \left(\frac{\tau_{1}}{\tau_{2}}\right)^{\es}\,,
\ee
so that 
\be
\cq=\sqrt{\frac{\pi}{2}}\frac{H(-\tau=x_{i}/k)}{x_{i}^{\es}\sqrt{2k^{3}}}
\ee
is independent of the choice of $x_{i}$ at the order we are working. On the other hand, the scale dependence is non-trivial: $\dep_{x=1} \propto\cq\propto k^{-3/2-\es}$. This constitutes one of the two sources of deviation from scale invariance. Our final goal is to compute the (gauge invariant) variable \cite{Weinberg:2008zzc}
\be\label{Rtof}
\R\equiv -\Psi-\frac{H}{\dot \phi}\dep\,,
\ee
with $g_{ij}=a^{2}\delta_{ij} \left(1-2\Psi\right)$. Outside of the horizon $\R$ has a constant and a decaying mode. To isolate the constant mode we first take the small $x$ limit of $\dep$
\be\label{vic2}
\dep\rightarrow -\frac{i}{\pi}2^{\nu}\Gamma(\nu)\cq \,x^{3/2+\es-\nu}\,.
\ee
Now we have to convert $\dep$ into $\R$ using $\dep=\R \sqrt{\epsilon}$, where $\epsilon$ depends slowly on time and gives the second and last contribution to deviations from scale invariance. For the purpose of computing deviations from scale invariance, the conversion can be done at any $x$, as long as this is the same for every mode so that we do not introduce any spurious scale dependence. This makes is clear that slow-roll corrections to $\nu$ in \eqref{vic2} only affect the amplitude, but not the scale dependence of the result since they do not affect any $k$-dependent quantity. In practice, the overall amplitude is captured most accurately if the conversion from $\dep$ to $\R$ is performed around horizon crossing for each mode, i.e.~$x=1$. Also, since $\cq$ does not depend on $x_{i}$ we can simply rewrite it evaluating $H$ at horizon crossing. The final result is then 
\be
|\R_{k}(x\rightarrow 0)|^{2}\simeq \frac{H^{2}}{4\epsilon k^{3}}=\frac{H^{2}_{\ast}}{4 \es k^{3}} \left(\frac{k}{k_{\ast}}\right)^{-2\es-\eta_{\ast}}\,,
\ee
where $-2\es$ and $\eta_{\ast}$ come from the scale dependence of $H^{2}$ and $1/\epsilon$, respectively. 

Summarizing, in perfect de Sitter $H$ is constant, every mode $\dep$ goes through the same history on the same background and the resulting spectrum for $\dep$ is scale invariant. In a quasi de Sitter expansion as relevant for inflation, $H$ and its derivatives vary slowly with time. As different modes go through the same history, they experience a slightly different background, due to the decrease in $H$ as inflation proceeds, and a slightly different factor $\sqrt{2\epsilon}$ for the conversion into $\R$. As long as we keep the momentum dependence of these two factors into account, we can ignore those slow-roll corrections, e.g.~in the last term of \eqref{eomfx}, which lead to only small corrections to the overall amplitude.

Let us now turn to the problem at hand
\be
\left[\frac{\partial^{2}}{\partial \tau^{2}}-\nabla^{2}+2aH\frac{\partial}{\partial\tau}+a^{2}V''\right]\dep=a^{2}\frac{\alpha}{f} \left( \edb-\ex{\edb}\right)\,.
\ee
or
\be
\left[\frac{\partial^{2}}{\partial \tau^{2}}+k^{2}+2aH\frac{\partial}{\partial\tau}+a^{2}V''\right]\dep_{k}=J_{\vec k}(\tau)\,,\label{eomf}\\
J_{\vec k}(\tau)\equiv a^{2} \frac{\alpha}{f}\int d^{3}x \,e^{-i\vec k\cdot \vec x}\,\edb(\vec x)\,,
\ee
Since we have already discussed the homogeneous part of the equation we know where the scale dependence will arise in solving the inhomogeneous part with the Green function method. Let us compute $\ex{\R^{\rm i.d.}_{\vec k}\R^{\rm i.d.}_{\vec k}}$ using
\be
\R^{\rm i.d.}_{\vec k}(\tau)=\frac{H(k)}{\dot \phi(k)}\int_{-\infty}^{0}d\tau'\,G_{k}(\tau,\tau')J_{\vec k}(\tau')\,,
\ee
where the Green function is given by
\be
G_{k}(\tau,\tau')=i\Theta(\tau-\tau')\left[\dep_{k}(\tau)\dep_{k}(\tau')-{\rm h.c.}\right]\,.
\ee
As discussed previously, $k$ dependence in $\dep$ comes just from $C_{k}$. Using this information and repeating the derivation of the power spectrum presented in \cite{BP}, one comes to the final result
\be\label{nsv}
\ex{\R_{\vec k}^{\rm i.d.}\R_{\vec k'}^{\rm i.d.}}=(2\pi)^{3}\delta^{3}(\vec k+\vec k') f_{2}(\xi(k))e^{4\pi \xi(k)}\frac{2\pi^{2}}{k^{3}} \left[\frac{H(k)^{2}}{(2\pi)^{2}}\frac{1}{2\es(k)}\right]^{2}\,.
\ee
where
\be\label{ff2}
f_{2}(\xi)&\equiv&\frac{\xi(k)^{2}}{8\pi}\int d^{3}q \left[1+\frac{q^{2}-\vec q\cdot \hat k}{q |\hat k-\vec q|}\right]^{2}\sqrt{q |\hat k\cdot \vec q|} \left[\sqrt{q}+\sqrt{|\hat k \cdot \vec q|}\right]^{2}\nonumber\,,\\
&&\qquad \times \mathcal{I}^{2}\left[\sqrt{8\xi}\left(\sqrt{q}+\sqrt{|\hat k\cdot \vec q|}\right)\right]
\label{f2full}
\ee
with
\be
\mathcal{I}(x)\equiv \sqrt{\frac{\pi}{2}}\int^{\infty}_{-k\tau} dx\,x^{3/2}\, {\rm Re} \left[H^{(1)}_{\nu}(x)\right] e^{-z\sqrt{x}}.
\ee
In \eqref{nsv}, $e^{4\pi \xi(k)}$ has been taken out of the integral appearing in the definition of $f_{2}$. This can be done because, as explained in \cite{BP}, the integral is supported around $|q|\sim|k|$. We can then rewrite the whole power spectrum, including both vacuum and inverse decay contributions as
\be
\langle \R(\vec k)\R(\vec k')\rangle&=&(2\pi)^{3} \delta \left(\vec k+\vec k'\right) \frac{2\pi^{2}}{k^{3}}\left[\frac{H^{2}_{\ast}}{2\pi|\dot \phi_{\ast}|}\right]^{2} \left(\frac{k}{k_{\ast}}\right)^{n_{s}-1}\\
&&\quad \times  \left[1+\left[\frac{H^{2}_{\ast}}{2\pi|\dot \phi_{\ast}|}\right]^{2} \left(\frac{k}{k_{\ast}}\right)^{n_{s}-1}\,f_{2}(\xi(k))\,e^{4\pi\xi(k)}\right]\,, 
\ee
with
\be 
n_s - 1 &=& -2\epsilon_* - \eta_* \simeq 6\epsilon_V + 2 \eta_V\,,\\
\xi (k)& =&\xi_{\ast} \left[1+\frac{\eta_{\ast}}{2}\log \left(\frac{k}{k_*}\right)\right]+\mathcal{O}(\epsilon^{2})\,.
\ee
Here the star in $H_{\ast}$ and in other scale dependent quantities indicates that these quantities are evaluated at horizon crossing of a pivot scale, which, for concreteness, we have chosen to be $k_{\ast}=0.002\, {\rm Mpc^{-1}}$.

We have one technical comment on computing $\mathcal{I}$ numerically. As explained in \cite{BPP}, we are only interested in super horizon modes, i.e.~modes with $-k\tau\ll 1$. Therefore, we can safely put the lower limit in the integral to 0. As we have reviewed, the slow-roll corrections to $\nu = 3/2+\mathcal{O(\epsilon,\eta)}$ do not affect the scale dependence of the correlation function, but just its amplitude. We can hence neglect them. The integral $\mathcal{I}$ can then be approximated to be 
\be
\mathcal{I}(z)&\simeq& \int_0^{\infty}dx \left(\sin x - x \cos x\right)e^{-z\sqrt{x}}\,,\\
&=&2+\frac{\sqrt{\pi}}{16}z\,{\rm Re} \left\{ e^{i(z^{2}-\pi)/4} \left(z^{2}-10i\right) {\rm Erfc} \left[e^{i\pi/4} \frac{z}{2} \right]+ \right.\nonumber\\
&&\hspace{3cm} \left. +e^{-i(z^{2}-\pi)/4} \left(z^{2}+10i\right) {\rm Erfc} \left[e^{-i\pi/4} \frac{z}{2}\right]\right\}\nonumber
\ee
where the second line is a semi-analytical expression useful in numerical computations. ${\rm Erfc}(x)\equiv 1-{\rm Erf}(x)$ is the complementary error function. 

\section{Technical details on the bispectrum constraint}\label{a:bi}

The three-point function is given by\cite{BP}
\be
\langle\prod_i^3 \mathcal{R}_{\vec{k}_i}(\tau)\rangle&=&  \int  \prod_i^3 \left[ \frac{d\tau_i}{a(\tau)}G_{k_{i}}(\tau,\tau_i)\frac{H(k_{i})}{\dot{\phi}(k_{i})} \right] \times \langle \prod^3_i  J_{\vec{ki}_i}(\tau_i) \rangle. 
\ee
It was shown in \cite{BPP} that this results in 
\be
\langle\prod_i^3 \mathcal{R}_{\vec{k}_i}(\tau)\rangle&=&\frac{\alpha^3 }{f^3 } \left[\prod_{i}^{3} \frac{H^{3}(k_{i})}{\dot \phi(k_{i}) k_i^2}\right] \delta(\sum \vec{k}_i) \int \frac{d^3 q_1}{(2\pi)^9}\prod_i^3 [\vec{\epsilon}((-1)^{i+1}\vec{q}_i)\cdot \vec{\epsilon}(\Theta(i)\vec{q}_{i+1})] \nonumber\\
&&\times \int_{-\infty}^0 d\tau_i[k_i\tau_i \cos (k_i\tau_i)-\sin (k_i \tau_i)] \times \mathcal{A}[\tau_i,|q_i|, |q_{i+1}|\,.
\label{bifull}
\ee
For convenience of notation we defined $\vec{q}_2 = \vec{k}_1 - \vec{q}_1$, $\vec{q}_3 = \vec{k}_3 + \vec{q}_1$, and $\vec{q}_4 =  \vec{q}_1$. In addition we defined a function $\Theta$ with the properties $\Theta(1)=1$, $\Theta(2)=\Theta(3) = -1$  while 
\be
\mathcal{A}(\tau, |\vec{a}|, |\vec{b}|) \equiv |\vec{a}| A_{,\tau}(\tau, |\vec{b}|)A(\tau, |\vec{a}|)+|\vec{b}| A_{,\tau}(\tau, |\vec{a}|)A(\tau, |\vec{b}|). 
\ee
In principle, $A=A_+$ is given by the Whittaker solution \eqref{Whittaker}. As was first argued in \cite{BPP} interesting physical effects occur in the interval $1/8\xi_*\lesssim -k\tau \lesssim 2\xi_*$. The solution for $A$ can then be approximated to be
\be
A(\tau,k) - \simeq \frac{1}{\sqrt{2k}}\left(\frac{-k\tau}{2\xi}\right)^{1/4}e^{\pi\xi-2\sqrt{-2\xi k\tau}}\,.
\ee 
Using this approximation we can compute the integrals over conformal time in \ref{bifull}, keeping track of all momentum dependence in $\xi$
\be
\prod_i^3\int_{-\infty}^0 d\tau_i[k_i\tau_i \cos (k_i\tau_i)-\sin (k_i \tau_i)] \times \mathcal{A}[\tau_i,|q_i|, |q_{i+1}|]&=& \frac{e^{6\pi\xi((\prod|\vec{q}_i|)^{1/3})}}{8} \times \;\;\;\;\;\;\;\;\;\;\;\;\;\;\;\;\;\;\;\;\;\;\;\;\;\;\;\;\;\;\;\;\;\;\;\;\ \nonumber
\ee
\be
&&\prod_i^3 \frac {|\vec{q}_i|^{1/2}}{k_i\sqrt{\xi(\vec{q_i})}} \left(\sqrt{|\vec{q}_i|\xi(|\vec{q}_{i+1}|)}+\sqrt{|\vec{q}_{i+1}|\xi(|\vec{q}_i|)}\right)\mathcal{I}\left( 2\left(\sqrt{2\xi(|\vec{q}_i|)\frac{|\vec{q}_i |}{|\vec{k}_i|}}+\sqrt{2\xi(|\vec{q}_{i+1}|)\frac{|\vec{q}_{i+1} |}{|\vec{k}_i|}}\right)\right)\nonumber\\
\label{integralex}
\ee
In section \ref{ss:bi} we argued that the $\xi$ in the exponential should dominate the scale dependence of the bispectrum. As argued in \cite{BP}, the support of the $q_{1}$ integral is for $q_{1}\sim k_{1}\sim k_{2}\sim k_{3}$, which explains why the resulting shape is close to equilateral. Hence we approximate $\xi$ in the exponential by
\be
6\pi\xi((\prod|\vec{q}_i|)^{1/3})\simeq6\pi\xi((\prod|\vec{k}_i|)^{1/3})\simeq \pi\xic \left[6+\eta_{\ast}\log \left(\frac{k_{1}k_{2}k_{3}}{k_{\ast}^{3}}\right)\right]\,.
\ee
We have checked explicitly that using the arithmetic instead of the geometric mean does not affect our final results. This is to be expected since we compare only shapes that peak in equilateral configurations, where the two means differ only very little. Then we can approximate the bispectrum as 
\be
\fid&=&\frac{\Delta_{\R,{\rm sr}}^{6}(k_{\ast})e^{6\pi\xic}f_{3}(\xic,1,1)}{\Delta_{\R}^{4}(k_{\ast})}\,,\\
F^{id}&=&\frac{3}{10}(2\pi)^{4} \Delta_{\R}^{4}(k_{\ast})\, \frac{\sum k_{i}^{3}}{\prod k_{i}^{3}}\,\frac{f_{3}(\xic,\frac{k_{2}}{k_{1}},\frac{k_{3}}{k_{1}})}{ f_{3}(\xic,1,1)}\left[\prod_{i}^{3} \left(\frac{k_{i}}{k_{\ast}}\right)^{\pi \xic \eta_{\ast}+n_{s}-1}\right]\,,\label{3pfall}
\ee
Defining $\vec{q}_i/k_1=\vec{Q}_i$ and $x_i = k_i/k_1$ we can write $f_3(\xi_*,k_2/k_1,k_3/k_1)$ as 
\be\label{f3x}
f_3(\xi_*, x_2, x_3)& = &\frac{5}{3\pi}\frac{1}{x_2 x_3 (1+x_2^3+x_3^3)}\int d^3Q_1 \prod_i^3 |\vec{Q}_i|^{1/2}\left(|\vec{Q}_i|^{1/2}+|\vec{Q}_{i+1}|^{1/2}\right)\nonumber\\
&&\mathcal{I}\left(2\sqrt{2\xi_*/x_i}(|\vec{Q}_i|^{1/2}+|\vec{Q}_{i+1}|^{1/2})\right)[\vec{\epsilon}((-1)^{i+1}\vec{Q}_i)\cdot \vec{\epsilon}(\Theta(i)\vec{Q}_{i+1})]\,.\quad
\label{f3final}
\ee
$f_3$ can be computed numerically by setting \cite{BPP}
\be
\vec{k}_1/k_1 = \hat{k}_1  = (1,0,0),
\ee
and
\be
\vec{k}_3/k_1 = -\frac{1}{2}\left(1-x_2^2+x_3^2,\sqrt{-(1-x_2+x_3)(1+x_2-x_3)(1+x_2+x_3)},0\right)\,,
\ee
and noting that to a generic vector $\vec{k} = k (\sin \theta \cos \phi,\sin \theta \sin \phi, \cos \theta)$ there exists a corresponding polarization vector 
\be
\vec{\epsilon}(\vec{k})=\frac{1}{\sqrt{2}}\left(\cos \theta \cos \phi-i\sin\phi,\cos \theta\sin \phi+i\cos \phi, -\sin\theta\right).
\ee
Using expressions \eqref{3pfall} and \eqref{f3final} and this parametrization we computed the fudge factor of \eqref{fudge}.

\section{Computation with the gauge fields}

Here is a list of useful formulae
\be
F^{i0}&=&a^{-2}(\dot A_{i}+\partial_{i}A^{0})\,,\\
F^{ij}&=&a^{-4}F_{ij}\,,\\
\dot A_{i}&=&a^{-1}\left(\Pi^{i}+()_{b}\right)-\partial_{i}A^{0}\,,\\
\partial_{t}(a^{3}F^{i0})&=& a \ddot A_{i}+aH\dot A_{i}+\partial_{t}(a\partial_{i}A^{0})\,,\\
\partial_{\mu}(a^{3}F^{\mu i})&=&-[a\ddot A_{i}-\frac{\partial_{k}^{2}A_{i}}{a}+aH\dot A_{i}+\partial_{t}(a\partial_{i}A^{0})]\,,\\
\partial_{\mu}(a^{3}\fb \tilde  F^{\mu i}/f)&=&\frac{\dot \fb}{f}\nabla\times\va\,,\\
\partial_{\mu}(a^{3}\vf \tilde  F_{b}^{\mu i}/f)&=&\frac{\dot \vf}{f}\nabla\times\ab-\frac1f (\nabla \vf)\times \dot{\vec{A_{b}}}\,,\\
\dot \Pi^{i}&=&aH\dot A_{i}+a\ddot A_{i}+\partial_{t}(a\partial_{i} A^{0}-\frac{\vf}{f}\nabla\times\ab-\frac{\fb}{f}\nabla\times \va)\,.
\ee
Using the Coulomb gauge $\nabla \cdot \va=0$ and
\be
\varepsilon_{ijk} \varepsilon^{imn}=\delta^{m}_j\delta^{n}_k - \delta^{n}_j\delta^{m}_k\,,
\ee 
one finds
\be
(\nabla\times \va)^{2}=\partial_{j}A_{k}\partial_{j}A_{k}\,.
\ee

%%%%%%%%%%%%%%%%%%%%%%%%%%%%%%%%%%%%%%%%%%%%%%%%%%%%%%%%

\end{document}